\documentclass[aps,prl,twocolumn,longbibliography,superscriptaddress,notitlepage,floatfix]{revtex4-2}
\renewcommand{\figurename}{Fig.}
\usepackage{amsmath,amsfonts,amssymb,color,epsfig,graphics,graphicx,latexsym,theorem,url,multirow,diagbox}
\usepackage[colorlinks,colorlinks,citecolor=blue,linkcolor=blue,urlcolor=blue]{hyperref}
\usepackage[utf8]{inputenc}
\usepackage{booktabs}
\usepackage[T1]{fontenc}
\usepackage{courier}
\usepackage{listings}
\usepackage{latexsym}
\usepackage{dcolumn}
\usepackage{comment}
\usepackage{times} 

\usepackage{algorithm}
\usepackage{algorithmic} 
\usepackage{braket}
\usepackage{bm}
\renewcommand{\algorithmicrequire}{\textbf{Input}}  
\renewcommand{\algorithmicensure}{\textbf{Output}} 

\definecolor{blue}{rgb}{0,0,1}
\definecolor{red}{rgb}{1,0,0}
\definecolor{green}{rgb}{0,1,0}

\begin{document}
\title{Quantum continual learning on a programmable
superconducting processor}
\affiliation{Zhejiang Key Laboratory of Micro-nano Quantum Chips and Quantum Control, School of Physics, Zhejiang University, Hangzhou 310027, China\\
$^2$ Center for Quantum Information, IIIS, Tsinghua University, Beijing 100084, China\\
$^{3}$ Shanghai Qi Zhi Institute, Shanghai 200232, China\\
$^4$ Institute of Automation, Chinese Academy of Sciences, Beijing 100084, China\\
$^{5}$ Instituut-Lorentz, Universiteit Leiden, P.O. Box 9506, 2300 RA Leiden, The Netherlands \\
$^{6}$ ZJU-Hangzhou Global Scientific and Technological Innovation Center, Hangzhou 310027, China\\
$^{7}$ International Research Centre MagTop, Institute of Physics, Polish Academy of Sciences, Aleja Lotnikow 32/46, PL-02668 Warsaw, Poland \\
$^{8}$ Hefei National Laboratory, Hefei 230088, China}

\author{Chuanyu Zhang$^{1}$}\thanks{These authors contributed equally to this work.}
\author{Zhide Lu$^{2,3}$}\thanks{These authors contributed equally to this work.}
\author{Liangtian Zhao$^{4}$}\thanks{These authors contributed equally to this work.}
\author{Shibo Xu$^{1}$}
\author{Weikang Li$^{2,5}$}
\author{Ke Wang$^{1}$}
\author{Jiachen Chen$^{1}$}
\author{Yaozu Wu$^{1}$}
\author{Feitong Jin$^{1}$}	
\author{Xuhao Zhu$^{1}$}
\author{Yu Gao$^{1}$}
\author{Ziqi Tan$^{1}$}
\author{Zhengyi Cui$^{1}$}
\author{Aosai Zhang$^{1}$}	
\author{Ning Wang$^{1}$}
\author{Yiren Zou$^{1}$}
\author{Tingting Li$^{1}$}
\author{Fanhao Shen$^{1}$}
\author{Jiarun Zhong$^{1}$}
\author{Zehang Bao$^{1}$}
\author{Zitian Zhu$^{1}$}
\author{Zixuan Song$^{6}$}
\author{Jinfeng Deng$^{1}$}
\author{Hang Dong$^{1}$}
\author{Pengfei Zhang$^{1,6}$}
\author{Wenjie Jiang$^{2}$}
\author{Zheng-Zhi Sun$^{2}$}
\author{Pei-Xin Shen$^{7}$}
\author{Hekang Li$^{6}$}
\author{Qiujiang Guo$^{1,6,8}$}
\author{Zhen Wang$^{1,8}$}
\author{Jie Hao$^{4}$}\email{jie.hao@ia.ac.cn}
\author{H. Wang$^{1,8}$}
\author{Dong-Ling Deng$^{2,3,8}$}\email{dldeng@tsinghua.edu.cn}
\author{Chao Song$^{1,8}$}\email{chaosong@zju.edu.cn}

\begin{abstract}
\textbf{
Quantum computers may outperform classical computers on machine learning tasks. In recent years, a variety of quantum algorithms promising unparalleled potential to enhance, speed up, or innovate machine learning have been proposed. Yet, quantum learning systems, similar to their classical counterparts, may likewise suffer from the catastrophic forgetting problem, where training a model with new tasks would result in a dramatic performance drop for the previously learned ones. This problem is widely believed to be a crucial obstacle to achieving continual learning of multiple sequential tasks. Here, we report an experimental demonstration of quantum continual learning on a fully programmable superconducting processor. In particular, we sequentially train a quantum classifier with three tasks, two about identifying real-life images and the other on classifying quantum states, and demonstrate its catastrophic forgetting through experimentally observed rapid performance drops for prior tasks. To overcome this dilemma, we exploit the elastic weight consolidation strategy and show that the quantum classifier can incrementally learn and retain knowledge across the three distinct tasks, with an average prediction accuracy exceeding \boldsymbol{$92.3\%$}. In addition, for sequential tasks involving quantum-engineered data, we demonstrate that the quantum classifier can achieve a better continual learning performance than a commonly used classical feedforward network with a comparable number of variational parameters. 
Our results establish a viable strategy for empowering quantum learning systems with desirable adaptability to multiple sequential tasks, marking an important primary experimental step towards the long-term goal of achieving quantum artificial general intelligence.
}
\end{abstract}

\maketitle


\noindent Continual learning, also known as incremental learning or lifelong learning, aims to empower artificial intelligence with strong adaptability to the non-stationary real world ~\cite{Wang2024Comprehensive,Ditzler2015Learning,chen2022lifelong}. 
It is a fundamental feature of natural intelligence, yet poses a notorious challenge for artificial intelligence based on deep neural networks.    
A major obstacle that hinders continual learning is catastrophic forgetting, where adaptation to a new task generally leads to a largely reduced performance on old tasks~\cite{McCloskey1989Catastrophic,Goodfellow2015Empirical}. This dilemma reflects a delicate trade-off between learning plasticity and memory stability: different sequential tasks correspond to different distributions and maintaining of plasticity would compromise stability in general~\cite{Wang2024Comprehensive}.  
In recent years, numerous efforts have been devoted to tackling this problem and the field of continual learning has been expanding rapidly~\cite{Wang2023Incorporating,vandeVen2022Three,Zeng2019Continual,Perkonigg2021Dynamic,Soltoggio2024Collective}, with the potential applications including medical diagnosis~\cite{Lee2020Clinical}, autonomous driving~\cite{Shaheen2022Continual}, and financial markets~\cite{Philps2019Continual}.

\begin{figure*}[t]
\centering
\includegraphics[width=0.93\textwidth]{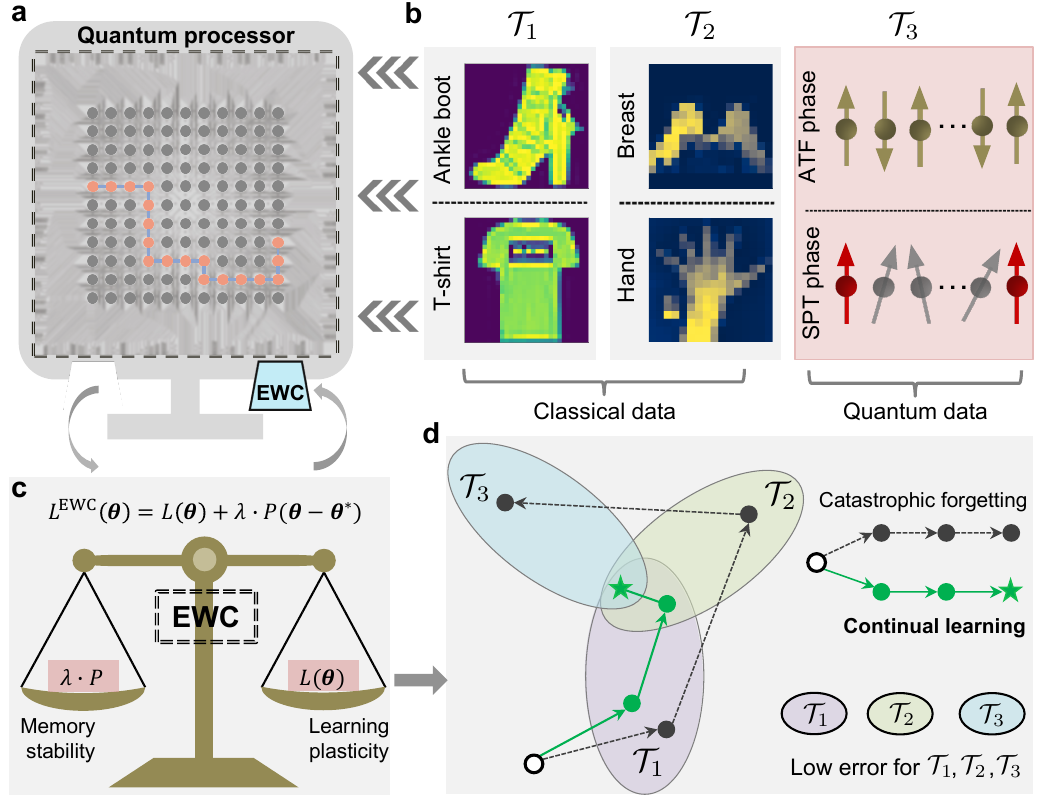}
\caption{
\textbf{Experimental quantum continual learning. a,} Exhibition of a $18$-qubit quantum classifier running on a superconducting processor. The used transmon qubits are marked in orange.
\textbf{b,} Training data for three consecutive learning tasks. 
$\mathcal{T}_1$ concerns the classification of images depicting ``T-shirt'' and ``ankle-boot'' sampled from the Fashion-MNIST dataset~\cite{Xiao2017FashionMNIST}.
$\mathcal{T}_2$  involves identifying images labeled as ``Hand''
and ``Breast'' from the magnetic resonance imaging dataset~\cite{Clark2013Cancer}. $\mathcal{T}_3$ is about recognizing quantum states in a symmetry-protected topological (SPT) phase and an antiferromagnetic (ATF) phase.
\textbf{c,} Illustration of elastic weight consolidation (EWC). EWC aims to balance memory stability for the previous task with learning plasticity for the new task.
The memory stability is preserved by penalizing the deviation of the parameter $\boldsymbol{\theta}$ from its optimal value $\boldsymbol{\theta}^{\star}$ for the previous task based on the importance of each parameter, which is measured by the Fisher information.
\textbf{d,} Conceptual diagram of catastrophic forgetting and continual learning. In a continual learning scenario, catastrophic forgetting refers to the dramatic performance drop on previous tasks after learning a new one. Continual learning is achieved when the learning system is able to maintain good performance on previous tasks while learning a new one.
}
\label{fig:illustration}
\end{figure*}

In parallel, the field of quantum computing has also made striking progress recently, with the experimental demonstration of quantum supremacy~\cite{Arute2019Quantum,Zhong2020Quantum,Wu2021Strong} and error correction codes~\cite{Acharya2023Suppressing,Bluvstein2024Logical,daSilva2024Demonstration,Acharya2024Quantum} marked as the latest breakthroughs. The interplay between quantum computing and machine learning gives rise to a new research frontier of quantum machine learning ~\cite{Biamonte2017Quantum,Dunjko2018Machine,Sarma2019Machine,Cerezo2022Challenges}. Different quantum learning algorithms~\cite{Harrow2009Quantum,Lloyd2014Quantum,Lloyd2018Quantum,Hu2019Quantum,Gao2018Quantum,Liu2021Rigorous} have been proposed and some of them haven been demonstrated in proof-of-principle experiments with current noisy intermediate-scale quantum (NISQ) devices~\cite{Preskill2018Quantum,Bharti2022Noisy}. However, to date most quantum learning models have been designed for a specific predefined task with a static data distribution and no experiment on quantum learning of multiple tasks sequentially has been reported. For quantum artificial intelligence systems to accommodate dynamic streams of data in the real world, the capability of continual learning is indispensable and crucial. 
To this end, a recent theoretical work has extended continual learning to the quantum domain~\cite{Jiang2022Quantum}. It is found that similar to classical learning models based on neural networks, quantum learning systems based on variational quantum circuits would suffer from catastrophic forgetting as well. In addition, a uniform strategy, namely the elastic weight consolidation (EWC) method~\cite{Kirkpatrick2017Overcoming}, has also been proposed to overcome this problem and achieve quantum continual learning. Despite this stimulating theoretical progress, experimental demonstration of quantum continual learning with NISQ devices is challenging and remains uncharted hitherto. To accomplish this, one faces at least two apparent difficulties:  (i) constructing an experimentally feasible quantum classifier with sufficient expressivity to accommodate multiple tasks with diverse non-stationary data distributions and (ii) designing a practical strategy to obtain Fisher information desired for implementing the EWC method, despite inevitable experimental noises.

In this paper, we overcome these difficulties and report the first experimental demonstration of quantum continual learning with a fully programmable 
superconducting quantum processor~(Fig.~\ref{fig:illustration}\textbf{a}).
We construct a quantum classifier with more than two hundred variational parameters, by using an array of $18$ transmon qubits featuring average simultaneous single- and two-qubit gate fidelities {greater than $99.96$\% and $99.68$\% respectively}. We demonstrate that, without EWC regulation, such a quantum classifier exhibits catastrophic forgetting when incrementally learning three tasks, including classifying real-life images and recognizing quantum phases~(Fig.~\ref{fig:illustration}\textbf{b}). However, by employing the EWC method, we can achieve a proper balance between memory stability for previous tasks and learning plasticity for new tasks, thus attaining quantum continual learning (Fig.~\ref{fig:illustration}\textbf{c}, \textbf{d}).
In addition, we compare the continual learning performance of quantum classifiers with that of classical classifiers in sequentially handling an engineered quantum task and a classical task.
We demonstrate that the quantum classifier can incrementally learn the two tasks with an overall accuracy up to $95.8\%$, exceeding the best overall accuracy of $81.3\%$ achieved by the classical classifier with a comparable number of parameters.  This manifests potential quantum advantages in continual learning scenarios.

\vspace{.5cm}
\noindent\textbf{\large{}Framework and experimental setup}

\noindent We first introduce the general framework for quantum continual learning \cite{Jiang2022Quantum}.
We consider a continual learning scenario involving three sequential tasks, denoted as $\mathcal{T}_k$ ($k=1,2,3$). As shown in Fig.~\ref{fig:illustration}(b), 
$\mathcal{T}_1$ concerns classifying clothing images labeled as ``T-shirts'' and ``ankle boot'' from the Fashion-MNIST dataset~\cite{Xiao2017FashionMNIST},
$\mathcal{T}_2$ concerns classifying medical magnetic resonance imaging (MRI) scans labeled as ``Hand'' and ``Breast''~\cite{Clark2013Cancer}, 
and $\mathcal{T}_3$ involves classifying quantum states in antiferromagnetic and symmetry-protected topological phases.
The learning process consists of three stages for sequentially learning these tasks. For the $k$-th task, we define the following cross-entropy loss function
\begin{equation}
\begin{aligned}
    L_k(\boldsymbol{\theta}) 
    &= \frac{1}{N_k} \sum_{i=1}^{N_k} L\left(h\left(\boldsymbol{x}_{k,i} ; \boldsymbol{\theta} \right), \mathbf{a}_{k,i}\right) \\
    &= - \frac{1}{N_k} \sum_{i=1}^{N_k} (\mathbf{a}_{k,i}^{0}\log \mathbf{g}_{k,i}^0 + \mathbf{a}_{k,i}^{1}\log \mathbf{g}_{k,i}^1),
    \label{eq:cross_entropy}
\end{aligned}
\end{equation}
where $N_k$ is the number of training samples, $\boldsymbol{x}_{k,i}$ denotes the $i$-th training sample, $\mathbf{a}_{k,i} = (\mathbf{a}_{k,i}^{0}, \mathbf{a}_{k,i}^{1})$ denotes the ground true label of $\mathbf{x}_{k,i}$ in the form of one-hot encoding, $h\left(\boldsymbol{x}_{k,i}; \boldsymbol{\theta} \right)$ denotes the hypothesis function for the quantum classifier parameterized by $\boldsymbol{\theta}$,
and $\mathbf{g}_{k,i} = (\mathbf{g}_{k,i}^{0}, \mathbf{g}_{k,i}^{1})$ denotes the probability of being assigned as label $0$ and label $1$ by the quantum classifier.
The performance of the quantum classifier is evaluated on the test dataset for $\mathcal{T}_k$. 
In our experiment, we first train the quantum classifier with the above loss function for each task sequentially. 
After each learning stage, the quantum classifier has a good performance on the current task but experiences a dramatic performance drop on the previous ones, which demonstrates the phenomena of catastrophic forgetting in quantum learning.

A salient strategy that can overcome catastrophic forgetting in quantum learning systems is the EWC method~\cite{Jiang2022Quantum,Kirkpatrick2017Overcoming}, which preserves memories for previous tasks by penalizing parameter changes according to the importance of each parameter. 
To demonstrate its effectiveness, in the $k$-th stage some regularization terms are added to the cross-entropy loss for $\mathcal{T}_k$, yielding a modified loss function
\begin{equation}
L_k^{\text{EWC}}(\boldsymbol{\theta})= L_k(\boldsymbol{\theta}) + \sum_{t=1}^{k-1} \frac{\lambda_{k,t}}{2} \sum_j F_{t,j}\left({\theta}_j-{\theta}^{\star}_{t,j} \right)^2,
\label{eq:loss_ewc}
\end{equation}
where $\lambda_{k,t}$ controls the regularization strength for $\mathcal{T}_t$ in the $k$-th stage; $\boldsymbol{\theta}^{\star}_{t}$ is the parameter obtained after the $t$-th stage;
$F_{t,j}$ denotes the Fisher information measuring the importance of the $j$-th parameter, which indicates how small changes to this parameter would affect the performance on $\mathcal{T}_{t}$ (Methods). 
A schematic illustration of the main idea for quantum continual learning is shown in Fig.~\ref{fig:illustration}{\textbf{c}, \textbf{d}}.

Our experiments are conducted on a flip-chip superconducting
quantum processor (Fig.~\ref{fig:illustration}\textbf{a}), which possesses $121$ transmon qubits arranged in a two-dimensional array with tunable nearest-neighbor couplings. We choose $18$ qubits (marked in orange in Fig.~\ref{fig:illustration}\textbf{a}) to implement a variational quantum classifier with a circuit depth of $20$ and $216$ trainable variational parameters~(Extended Data Fig.~\ref{fig:18_qubit_circuit}). To achieve a better learning performance, we push the average simultaneous two-qubit gate fidelities greater than $99.68\%$ through optimizing the device fabrication and control processes. We mention that the gradients and Fisher information desired in updating the quantum classifier are obtained by measuring observables directly in the experiment based on the ``parameter shift rule''~\cite{Mitarai2018Quantum}. Supplementary Section IIA provides more details about the characterization of the device.

\vspace{.5cm}
\noindent\textbf{\large{}Demonstration of catastrophic forgetting}

\noindent To demonstrate catastrophic forgetting in quantum learning, we train sequentially the quantum classifier with the loss function defined in Equation~(\ref{eq:cross_entropy}) for the three tasks. Our experimental results are displayed in Fig.~\ref{fig:exp_results_qcl}\textbf{a}.
The learning process comprises three stages.
In the first stage, the quantum classifier is trained to learn $\mathcal{T}_1$. After $20$ epochs of parameter updating, the prediction accuracy for classifying clothing images reaches $99\%$.
In the second stage, the quantum classifier is retrained on the training data for $\mathcal{T}_2$.
After $28$ epochs, it attains a classification accuracy of $99\%$ on $\mathcal{T}_2$.
However, after this training stage the performance on $\mathcal{T}_1$ drops dramatically to $54\%$. 
In the third stage, the quantum classifier is further trained to recognize quantum phases. 
After $18$ epochs, the quantum classifier achieves an accuracy of $100\%$. However, the accuracy for $\mathcal{T}_2$  and $\mathcal{T}_1$ dramatically falls to $64\%$ and $55\%$. 
These experimental results clearly showcase the phenomena of catastrophic forgetting in quantum learning.

\vspace{.5cm}
\noindent\textbf{\large{}Continual learning with EWC}

\noindent In this section, we show that the above demonstrated catastrophic forgetting can be effectively overcome with the EWC method. To this end, we train sequentially the quantum classifier with the modified loss function that includes the EWC regularization as defined in Equation~(\ref{eq:loss_ewc}). Our experimental results are shown in Fig.~\ref{fig:exp_results_qcl}\textbf{b}.
We observe that after the second learning stage, the prediction accuracy for $\mathcal{T}_2$  reaches $95\%$ while the accuracy for $\mathcal{T}_1$ still maintains $97\%$. After the third learning stage, the prediction accuracy for $\mathcal{T}_3$ reaches $96\%$, while it remains $88\%$ and $93\%$ for $\mathcal{T}_2$ and $\mathcal{T}_1$, respectively. This is in sharp contrast to the case without the EWC strategy, where it drops to $64\%$ and $55\%$, respectively. 
After training, we plot the distribution of the experimentally measured $\langle\hat{\sigma}^{z}_9\rangle$, whose sign determines the assigned labels~(Methods), for all test data samples, as shown in Fig.~\ref{fig:exp_results_qcl}\textbf{c}. 
It is clear that when applying EWC, data samples from $\mathcal{T}_1$ and $\mathcal{T}_2$  with different labels are far more distinguishable than the case without EWC, which confirms that the learned knowledge for $\mathcal{T}_1$ and $\mathcal{T}_2$ are effectively preserved with EWC.
\begin{figure*}[t]
\centering
\includegraphics[width=0.98\textwidth]{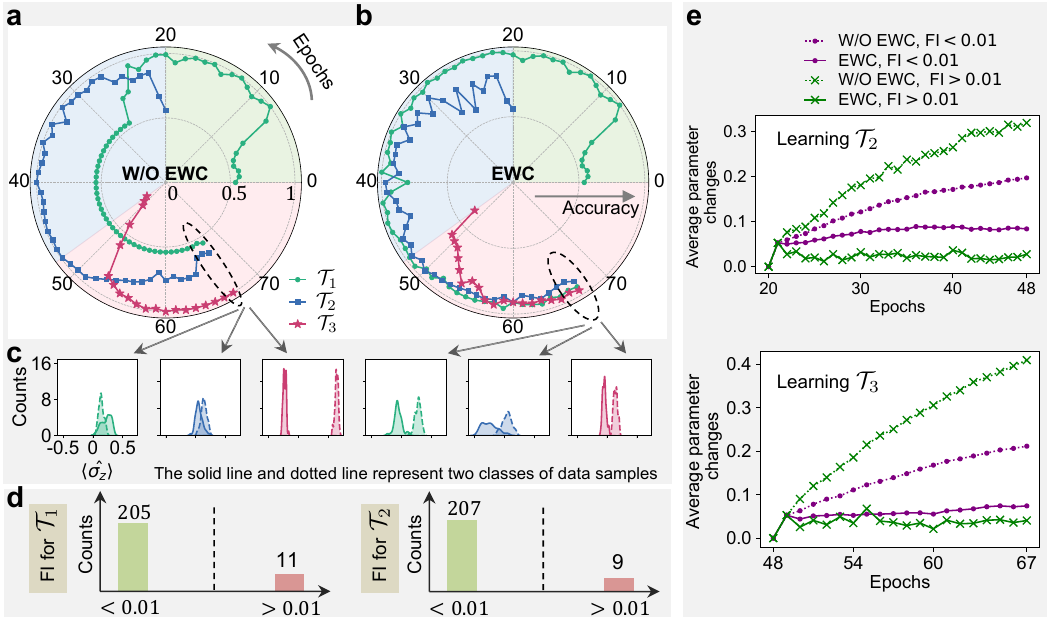}
\caption{
\textbf{Experimental results for continually learning three tasks.}
\textbf{a and b}, The prediction accuracy for three sequential tasks at each epoch during the continual learning process of the quantum classifier. 
Tasks $\mathcal{T}_1$, $\mathcal{T}_2$, and $\mathcal{T}_3$ are marked in green, blue, and red, respectively.
The right (left) figure shows the case where EWC (no EWC) is employed.
\textbf{c}, Distribution of the experimentally measured expected values $\langle\hat{\sigma}^{z}_9\rangle$, which determine the prediction label of input data, for all test samples after training. For each task, the solid line and dotted line correspond to two classes of data samples, respectively. A greater separation between the two distributions means better classification performance.
\textbf{d}, Distribution of Fisher information (FI) for all parameters after learning each task.
\textbf{e}, Average parameter change compared to the obtained parameters for previous tasks during the learning stage for the new task.
The top (bottom) figure corresponds to the learning for $\mathcal{T}_2$  ($\mathcal{T}_3$).
}
\label{fig:exp_results_qcl}
\end{figure*}

\begin{figure*}[t]
\centering
\includegraphics[width=0.95\textwidth]{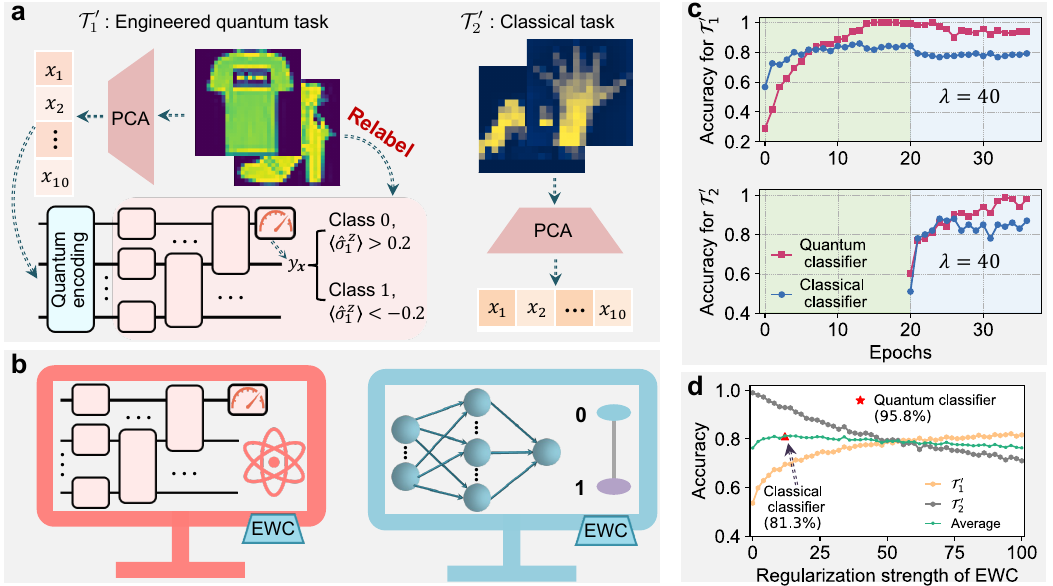}
\caption{
\textbf{Comparison between the continual learning performances for the quantum and classical classifiers.} 
\textbf{a}, Training data for two sequential tasks $\mathcal{T}_1^{\prime}$ and $\mathcal{T}_2^{\prime}$ and.
For $\mathcal{T}_1^{\prime}$, we choose clothing images as the source data and use principle component analysis (PCA) to reduce the dimension of each image to obtain a ten-dimensional vector. The ground truth label of each input sample is determined by a local observable $\langle\hat{\sigma}^{z}_{1}\rangle$ evolved under a quantum circuit with randomly chosen gate parameters.
For $\mathcal{T}_2^{\prime}$, we choose medical images as the source data. We use PCA to compress each image to a ten-dimensional vector as the input data. The label of the input vector is determined by the category of its original images.
\textbf{b}, Schematic illustration of a quantum classifier and a classical classifier based on feedforward neural network (FFNN).
\textbf{c}, Prediction accuracy for two sequential tasks as functions of training epochs during the continual learning process. For both quantum and classical classifiers, EWC is employed with the regularization strength set as $40$.
\textbf{d}, Continual learning performance of the classical classifier as a function of regularization strength.
For the classical classifier based on FFNN, we employ EWC with different regularization strengths.
For each regularization strength, we train the classical classifier for $50$ times and plot the mean prediction accuracy for $\mathcal{T}_1^{\prime}$ and $\mathcal{T}_2^{\prime}$, and their averages.
The optimal achievable overall performance, evaluated as the average of the accuracy on $\mathcal{T}_1^{\prime}$ and $\mathcal{T}_2^{\prime}$, is $81.3\%$ for the classical classifier and $95.8\%$ for the quantum classifier.
}
\label{fig:quantum_advantages}
\end{figure*}

To further understand how EWC balances the stability-plasticity trade-off for quantum classifiers, we analyze the average parameter changes in cases with EWC. 
According to Equation~(\ref{eq:loss_ewc}), for parameters with larger Fisher information, their deviations from the optimal values for previous tasks will cause a relatively more significant increase in the loss function. Therefore, the parameters with large Fisher information tend to undergo only small adjustments when learning the new task, so as to minimize the increase in the loss function. To verify this understanding experimentally, we measure $F_{1,i}$ for each parameter after the first learning stage. As shown in Fig.~\ref{fig:exp_results_qcl}\textbf{d}, we find that only $11$ parameters have $F_{1,i}$ values larger than $0.01$, while the other $205$ parameters have $F_{1,i}$ values less than $0.01$. 
Based on this, we divide all parameters into two groups and plot the average parameter change for each group during the second learning stage for $\mathcal{T}_2$.
The results are shown in Fig.~\ref{fig:exp_results_qcl}\textbf{e}. From this figure, it is clear that in the case with EWC, the parameters with large Fisher information ($>0.01$) experience smaller changes on average than the parameters with small Fisher information ($<0.01$).
This is consistent with the goal of EWC, which is to ensure that more important parameters experience smaller changes, therefore better maintaining the performance on $\mathcal{T}_1$.
The average parameter change in the third stage for learning $\mathcal{T}_3$ is also plotted in Fig.~\ref{fig:exp_results_qcl}\textbf{e}, which shows a similar observation.
Compared to the case without EWC, parameters with both large and small Fisher information exhibit smaller changes.
This is consistent with the fact that the added regularization terms will in general constrain the change of parameters. 
These experimental results unambiguously demonstrate the effectiveness of EWC in mitigating catastrophic forgetting in quantum continual learning scenarios.

We remark that, after learning each task, only a small portion of all parameters have relatively large Fisher information. This reflects that memories for the task can be preserved by selectively stabilizing these parameters. The majority of parameters, with relatively small Fisher information, retain a relatively large space to learn new tasks in subsequent stages.
This selective stabilization mechanism in EWC mirrors biological learning processes, where old memories are preserved by strengthening previously learned synaptic changes~\cite{Wang2023Incorporating}. We also mention that, although various continual learning strategies other than the EWC method exist \cite{Wang2024Comprehensive}, overcoming the catastrophic forgetting problem has been proved to be NP-hard in general~\cite{Knoblauch2020Optimal}. As a result, we do not expect that the EWC method for quantum continual learning demonstrated above to be universally applicable to arbitrary sequential tasks or have the optimal performance on given tasks.

\vspace{.5cm}
\noindent\textbf{\large{}Potential quantum advantages}

\noindent We consider two classification tasks with distinct data distributions: an engineered quantum task denoted as $\mathcal{T}^{\prime}_1$ and a classical task denoted as $\mathcal{T}^{\prime}_2$.
As shown in Fig.~\ref{fig:quantum_advantages}\textbf{a}, $\mathcal{T}^{\prime}_1$ involves classifying engineered training data samples with target functions generated by a quantum model \cite{Havlicek2019Supervised,Huang2021Power,Jerbi2023Quantum}, whereas $\mathcal{T}^{\prime}_2$ involves identifying medical images.
To construct the dataset for $\mathcal{T}^{\prime}_1$, we choose clothing images of ``T-shirt'' and ``ankle boot'' as the source data and use principle component analysis (PCA) to compress the dimension of each image to ten.
We generate the ground-truth label of each input data using a quantum model. To realize this, the ten-dimensional vector of each data is first encoded as a ten-qubit quantum state. The ground truth label is then taken as a local observable $\langle\hat{\sigma}^{z}_{1}\rangle$ evolved under a given quantum circuit with randomly chosen variational parameters (Methods).
For $\mathcal{T}^{\prime}_2$, we use medical images as the source data. We similarly use PCA to compress each image to a ten-dimensional vector. The ground truth label of each data sample is its original label ``hand'' or ``breast''.

In a continual learning scenario involving these two tasks in sequence, we compare the performance of quantum and classical models.
For quantum learning, we experimentally implement a ten-qubit quantum circuit classifier with a total of $90$ variational parameters (Fig.~\ref{fig:quantum_advantages}\textbf{b}, left). 
The learning process consists of two stages. In each stage, the ten-dimensional vector of each input data is embedded as a ten-qubit quantum state, followed by the ten-qubit variational quantum classifier~(Methods).
In Fig.~\ref{fig:quantum_advantages}\textbf{c}, we present the experimental results.
In the first stage, the quantum classifier is trained on $\mathcal{T}^{\prime}_1$, achieving $99.1\%$ prediction accuracy after $20$ epochs of parameter updating.
In the second stage of learning $\mathcal{T}^{\prime}_2$, the EWC method is employed with a regularization strength of $\lambda_q=40$.
After $16$ training epochs, the accuracy on $\mathcal{T}^{\prime}_2$ reaches $98\%$, while the accuracy on $\mathcal{T}^{\prime}_1$ slightly drops to $93.7\%$. The overall performance, typically evaluated by the average accuracy of the two tasks, is $95.8\%$.

For classical learning, we use a three-layer feedforward neural network with $241$ variational parameters as the classical classifier (Fig.~\ref{fig:quantum_advantages}\textbf{b}, right). In each learning stage, the ten-dimensional vector is directly taken as the input data of the classical classifier~(Methods).
We present the numerical results in Fig.~\ref{fig:quantum_advantages}\textbf{d}. We find that the classical classifier struggles to achieve good performance on both tasks simultaneously, as $\mathcal{T}^{\prime}_1$ and $\mathcal{T}^{\prime}_2$ largely interfere with each other. The dominance of each task depends on the regularization strength $\lambda_c$ used in EWC.
For small values of $\lambda_c$, the classical classifier achieves high accuracy on $\mathcal{T}^{\prime}_2$ but performs poorly on $\mathcal{T}^{\prime}_1$, indicating catastrophic forgetting.
As $\lambda_c$ increases, the classical classifier places more weight on preserving old memories for $\mathcal{T}^{\prime}_1$. This leads to an improvement in performance on $\mathcal{T}^{\prime}_1$ and a drop in performance on $\mathcal{T}^{\prime}_2$.
When $\lambda_c$ is increased to a large value ($\lambda_c=100$), the classical classifier almost completely loses its learning plasticity for $\mathcal{T}^{\prime}_2$ in the second learning stage.
The best overall performance that can be achieved by the classical classifier is $81.3\%$.

The comparison between quantum and classical models shows that quantum models can outperform classical models in certain continual learning scenarios, despite containing fewer variational parameters. This agrees with the theoretical predictions that quantum neural networks in general possess larger expressive power~\cite{Du2020Expressive} and effective dimension~\cite{Abbas2021Power} than classical ones with a comparable number of parameters, thus would better accommodate distribution differences among multiple tasks and lead to superior overall performance in continual learning scenarios.

\vspace{.5cm}
\noindent\textbf{\large{}Discussion and outlook}

\noindent 

\noindent In classical continual learning, a variety of strategies other than the EWC method, such as orthogonal gradient projection~\cite{Farajtabar2020Orthogonal} and parameter allocation~\cite{Mallya2018Piggyback}, have been proposed to overcome catastrophic forgetting. These strategies might also be adapted to quantum continual learning scenarios and their experimental demonstrations would be interesting and important. Along this direction, it is, however, worthwhile to mention a subtle distinction between quantum and classical continual learning. In the quantum domain, due to the no-cloning theorem~\cite{Wootters1982Single} and the difficulty in building long-live quantum memories~\cite{Terhal2015Quantum}, one cannot duplicate unknown quantum data and store them for a long time. As a result, replay-based strategies relying on recovering the old data distributions~\cite{Isele2018Selective,Rolnick2019Experience} might be inaccessible in the case of quantum continual learning. 
In addition, this work primarily focuses on classification tasks in the framework of supervised learning. The extension of quantum continual learning to unsupervised and reinforcement learning presents more technical difficulties and has yet to be achieved in both theory and experiment. 

Enabling a quantum artificial intelligence system to accommodate a dynamic stream of tasks and demonstrate its advantage in real-life applications remains largely unclear and demands long-term research. 
Our work makes a primary step in this direction by demonstrating the issue of catastrophic forgetting and the effectiveness of the EWC method for quantum continual learning in experiments, thus providing a valuable guide towards achieving quantum artificial general intelligence in the future. 

\vspace{.6cm}
\noindent\textbf{\large{}Data availability}\\
The data presented in the figures and that support the other findings of this study will be made publicly available for download on Zenodo/Figshare/Github upon publication.

\vspace{.5cm}
\noindent\textbf{Acknowledgement} 
We thank J. Eisert, M. Hafezi, D. Yuan, and S. Jiang for helpful discussions. 
{The device was fabricated at the Micro-Nano Fabrication Center of Zhejiang University.  We acknowledge support from the Innovation Program for Quantum Science and Technology (Grant Nos.~2021ZD0300200 and 2021ZD0302203), the National Natural Science Foundation of China (Grant Nos.~12174342, 92365301, 12274367, 12322414, 12274368, 12075128, and T2225008), the National Key R\&D Program of China (Grant No.~2023YFB4502600), and the Zhejiang Provincial Natural Science Foundation of China (Grant Nos.~LDQ23A040001, LR24A040002). Z.L., W.L., W.J., Z.-Z.S., and D.-L.D. are supported in addition by Tsinghua University Dushi Program, and the Shanghai Qi Zhi Institute.
P.-X.S. acknowledges support from the European Union's Horizon Europe research and innovation programme under the Marie Skłodowska-Curie Grant Agreement No.~101180589 (SymPhysAI) and the National Science Centre (Poland) OPUS Grant No.~2021/41/B/ST3/04475.
Views and opinions expressed are however those of the author(s) only and do not necessarily reflect those of the European Union or the European Research Executive Agency. Neither the European Union nor the granting authority can be held responsible for them.
}

\vspace{.7cm}
\noindent\textbf{\large{}Methods}

\noindent\textbf{Variational quantum classifiers} \\
We build the quantum classifiers with multiple blocks of operations, as illustrated in Extended Data Fig.~\ref{fig:18_qubit_circuit} and Extended Data Fig.~\ref{fig:exp_relabel}.
Each block contains three layers of single-qubit gates with programmable rotation angles and ends with two layers of entangling gates for leveraging the exponentially large Hilbert space and establishing quantum correlations among the qubits.
For classification tasks, the quantum classifier assigns a label to each input data based on the measured expectation value of the Pauli-Z operator on the $m$-th qubit, $\langle\hat{\sigma}^z_{m}\rangle$: a label for one class is assigned when $\langle\hat{\sigma}^z_m\rangle \geq0$, while a label for the other class is assigned when $\langle\hat{\sigma}^z_m\rangle <0$.
In the experiment for learning $\mathcal{T}_1$, $\mathcal{T}_2$ and $\mathcal{T}_3$, we use $18$ qubits with four blocks to construct the quantum classifier with a total of $216$ variational parameters, where the entangling gates are selected as CNOT gates, and $m=9$.
In the experiment for learning $\mathcal{T}^{\prime}_1$ and $\mathcal{T}^{\prime}_2$, we construct a ten-qubit quantum classifier with three blocks containing a total of $90$ variational parameters, where the entangling gates are selected as CZ gates, and $m=1$.

\vspace{.3cm}
\noindent\textbf{Dataset generation} \\
The datasets for $\mathcal{T}_1$ and $\mathcal{T}_2$ are composed of images randomly selected from the Fashion-MNIST dataset~\cite{Xiao2017FashionMNIST} and the MRI dataset~\cite{Clark2013Cancer}, respectively. 
The quantum dataset for $\mathcal{T}_3$ is composed of ground states of the cluster-Ising Hamiltonian~\cite{Smacchia2011Statistical} in the ATF and SPT phases. 
We prepare approximate ground states in our experiments by executing a variational circuit. We first train the variational circuit on a classical computer with the aim of minimizing the energy expectation value for the output states. 
We then experimentally implement the variational circuit using the parameters obtained in the classical simulation. 
To characterize our quantum state preparation, we measure the string order parameter for these prepared states. In Supplementary
Sec.~IIB, we provide a detailed discussion about the quantum state preparation.
For each of $\mathcal{T}_1$, $\mathcal{T}_2$ and $\mathcal{T}_3$, we construct a training set with $500$ data samples and a test set with $100$ data samples.

To construct the dataset for $\mathcal{T}^{\prime}_1$, we use the input data sourced from the Fashion-MNIST dataset. Specifically, we randomly select $1200$ images labeled as ``T-shirt'' and ``ankle boot''. 
We first perform PCA to compress these images to ten-dimensional vectors.
Subsequently, each feature of these ten-dimensional vectors is further normalized to have a mean value of $0$ and a standard deviation of $1$. 
As depicted in Extended Data Fig.~\ref{fig:exp_relabel}, we generate the label $g(\boldsymbol{x})$ for each data sample $\boldsymbol{x}$ using functions generated by a quantum model.
To this end, we first use the feature encoding proposed in Ref.~\cite{Havlicek2019Supervised} to encode $\boldsymbol{x}$ into a quantum state. 
We show the quantum circuit for the feature encoding in Supplementary Fig.~2\textbf{b}.
We then experimentally implement the quantum circuit model with three blocks of operations. 
The variational parameters $\boldsymbol{\theta}$ for the circuit are randomly generated within the range of $[0,2\pi]^{90}$.
The ground true label $g(\boldsymbol{x})$ is determined by the local observable $\langle \hat{\sigma}^z_1 \rangle$ evolved under the above circuit model: $g(\boldsymbol{x}) = 0$ if $\langle \hat{\sigma}^z_1 \rangle > 0.2$ and $g(\boldsymbol{x}) = 1$ if $\langle \hat{\sigma}^z_1 \rangle < -0.2$.
In our experiment, we obtain a total of $667$ data samples with $g(\boldsymbol{x})$ being $0$ and $1$. We select $556$ of them as the training dataset and the other $111$ of them as the test dataset.

To construct the dataset for $\mathcal{T}^{\prime}_2$, we use the data from the MRI dataset. We randomly select $600$ images labeled as ``hand'' and ``breast''. We also employ PCA to compress these images to ten-dimensional vectors. 
The ground true label of each ten-dimensional vector is just the label of the corresponding original image. We divide $600$ samples into a training dataset of size $500$ and a test dataset of size $100$.

\vspace{.3cm}
\noindent\textbf{Data encoding} \\
In our experiments, we utilize different strategies to encode different types of data. 
We use the interleaved block encoding strategy~\cite{Ren2022Experimental} to encode classical images in the dataset for $\mathcal{T}_1$ and $\mathcal{T}_2$.
For each classical image, we first reduce its size to $16\times16$ grayscale pixels and flatten it into a $256$-dimensional vector. We then normalize the vector and add up the adjacent entries to obtain a $128$-dimensional vector $\boldsymbol{x}$.
As shown in Extended Data Fig.~\ref{fig:18_qubit_circuit},
we assign each single-qubit rotation gates with an angle of $2x_i+\theta_i$, where $\theta_i$ is a variational parameter. We choose $128$ rotation gates and assign the corresponding $x_i$ with an entry of $\boldsymbol{x}$. For the remaining $88$ rotation gates, we set the corresponding $x_i$ with $0$ values.
For the quantum data in $\mathcal{T}_3$, the quantum classifier can naturally handle these quantum states as input after their preparation on quantum devices.

For $\mathcal{T}^{\prime}_1$, we encode the data into the quantum classifier using the feature encoding with the circuit structure shown in Supplementary Fig.~2\textbf{b}.
For $\mathcal{T}^{\prime}_2$, we use the rotation encoding to encode the data, with the circuit structure depicted in Supplementary Fig.~2\textbf{c}.

\vspace{.3cm}
\noindent\textbf{Gradients and Fisher information} \\
We minimize the loss function in Equation~(\ref{eq:cross_entropy}) by adapting the gradient descent method.
Based on the chain rule, the derivatives of $L$ with respect to the $j$-th parameter $\theta_j$ can be expressed as:
\begin{equation}
    \frac{\partial L\left(h\left(\boldsymbol{x}_{k,i} ; \boldsymbol{\theta} \right), \mathbf{a}_{k,i}\right)}
    {\partial \theta_j} = 
    -\frac{\mathbf{a}_{k,i}^0}{\mathbf{g}_{k,i}^0} \frac{\partial \mathbf{g}_{k,i}^0}{\partial \theta_j}-\frac{\mathbf{a}_{k,i}^1}{\mathbf{g}_{k,i}^1} \frac{\partial \mathbf{g}_{k,i}^1}{\partial \theta_j}  .
\end{equation}
In our experiment, $\mathbf{g}_{k,i}^0$ and $\mathbf{g}_{k,i}^1$ are determined by the local observable $|0\rangle\langle0|_m$ and $|1\rangle\langle1|_m$ on the $m$-th qubit, respectively.
As all variational parameters in the quantum classifier take the form of $\exp(-\frac{i}{2}\theta P_n)$ ($P_n$ belongs to the Pauli group), the derivatives of $\mathbf{g}_{k,i}^l$ can be computed via the ``parameter-shift rule''~\cite{Li2017Hybrid,Mitarai2018Quantum}:
\begin{equation}
    \frac{\partial \mathbf{g}_{k,i}^{l}}{\partial \theta_j} = \frac{(\mathbf{g}_{k,i}^{l})^{+} - (\mathbf{g}_{k,i}^{l})^{-}}{2} ,
\end{equation}
where $l=0, 1$, and $(\mathbf{g}_{k,i}^{l})^{\pm}$ denotes the expectation values of the local observables with parameter $\theta_j$ being $\theta_j \pm \frac{\pi}{2}$.
We directly measure $(\mathbf{g}_{k,i}^{l})^{\pm}$ in experiments to obtain the quantum gradients, based on which we adapt the gradient descent method assisted by the Nadam optimizer~\cite{Dozat2016Incorporating} to optimize the quantum classifier. The learning rate is set as $0.05$ in experiments.

After learning the $k$-th task, we need to obtain the Fisher information $F_{k,j}$ for measuring the importance of each variational parameter $\theta_j$.
Based on the derivatives of the loss function at $\boldsymbol{\theta}_k^{\star}$, we estimate $F_{k,j}$ as:
\begin{equation}
    F_{k,j} = \frac{1}{N_k}
    \sum_{i=1}^{N_k}
    \left(\left.\frac{\partial L\left(h\left(\boldsymbol{x}_{k,i} ; \boldsymbol{\theta} \right), \mathbf{a}_{k,i}\right)}
    {\partial \theta_j}\right|_{\boldsymbol{\theta}=\boldsymbol{\theta}_k^{\star}}\right)^2  ,
    \label{eq:fisher}
\end{equation}
The notations here follow those in Equation~(\ref{eq:cross_entropy}) and Equation~(\ref{eq:loss_ewc}).
The detailed derivation of $F_{k,j}$ is provided in Supplementary Sec.~I.B.

\vspace{.3cm}
\noindent\textbf{Training with EWC} \\
To sequentially learn multiple tasks without catastrophic forgetting, we adapt the EWC method. 
The learning process consists of multiple stages. Initially, each variational parameter in the quantum classifier is randomly chosen within the range of $[-\pi, \pi]$.
In the $k$-th stage, the quantum classifier is trained with the modified loss function $L^{\text{EWC}}$ as defined in Equation~(\ref{eq:loss_ewc}). 
At each training epoch, we calculate the gradients of $L^{\text{EWC}}$ on $25$ data samples randomly selected from the training dataset, and evaluate the learning performance on all data samples in the test dataset.
After the $k$-th stage, we obtain the Fisher information $F_{k,i}$ for all variational parameters, which is used in the subsequent learning stage~(see Supplementary Sec.~IB for detailed algorithms).
In the experiment for learning $\mathcal{T}_1$, $\mathcal{T}_2$ and $\mathcal{T}_3$, we set $\lambda_{2,1}=60$ in the second stage, and $\lambda_{3,1}=0$ and $\lambda_{3,2}=60$ in the third stage. We cancel the regularization term for $\mathcal{T}_1$ ($\lambda_{3,1}=0$) in the third stage for two reasons. First, we expect the model to have fewer restrictions and thus more flexibility in adjusting parameters to learn $\mathcal{T}_3$. 
Second, after the second stage, the obtained parameters $\boldsymbol{\theta}^{\star}_{2}$ can maintain knowledge for $\mathcal{T}_1$ since the regularization term for $\mathcal{T}_1$ is added during the second stage. 
Thus, by only adding the regularization term for $\mathcal{T}_2$, we can still preserve the learned knowledge from $\mathcal{T}_1$, as evidenced by the experimental results.
Although the information from $\mathcal{T}_1$ will decay as we sequentially learn more tasks, considering there are only three tasks in total, it is reasonable to set $\lambda_{3,1}=0$ for simplicity.

\vspace{.1cm}
\noindent\textbf{Classical learning models for comparison} \\
We specify classical feedforward networks used in numerical simulations for comparison.
For quantum learning, we use the quantum circuit classifier belonging to the same variational family as those employed for relabeling data for $\mathcal{T}^{\prime}_1$, but initialized with different parameters randomly generated from $[0,2\pi]^{90}$.
This quantum classifier contains a total of $90$ variational parameters.
For classical learning, we use a three-layer FFNN with ten neurons in the input layer, $20$ neurons in the hidden layer, and one neuron in the output layer. The activation function used is the sigmoid function.
This FFNN contains a total of $241$ variational parameters.
The ten neurons in the input layer encode the ten-dimensional input data vectors for the two tasks.
The neuron in the output layer determines the prediction outcome for the input data: if the output value is greater than $0.5$, the input data is assigned to class $0$; if the output value is less than $0.5$, it is assigned to class $1$.

\vspace{.5cm}
\noindent\textbf{Extended Data} 
\renewcommand{\figurename}{Extended Data Fig.}
\setcounter{figure}{0}

\begin{figure*}[htb]
\includegraphics[width=0.85\textwidth]{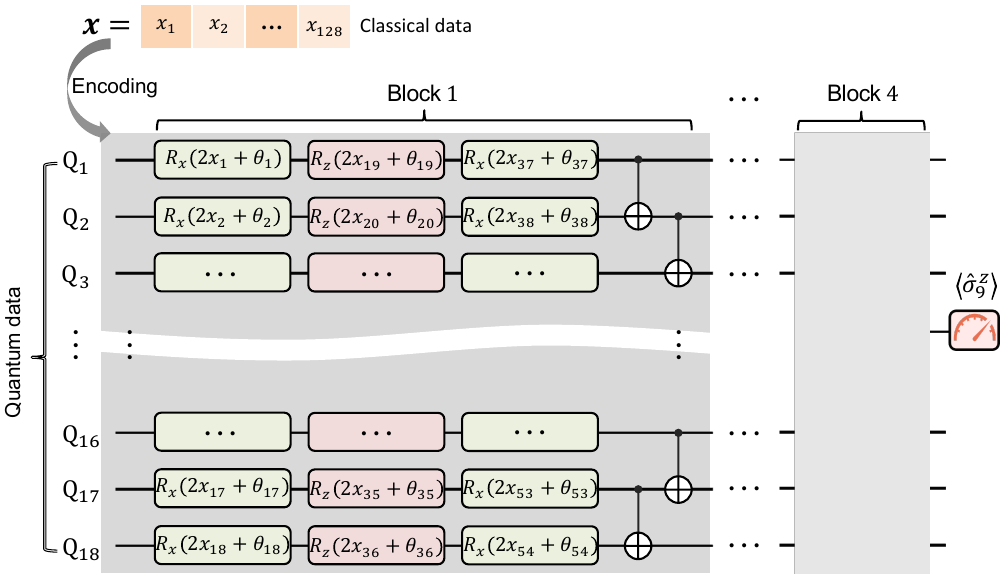}
\caption{ 
\textbf{Quantum circuit classifier with $18$ superconducting qubits for learning three sequential tasks.}
The circuit consists of four blocks of operations with a total of $216$ variational parameters. Each block performs three consecutive single-qubit rotation gates on all qubits, followed by two layers of CNOT gates applied to adjacent qubits.
The quantum classifier adapts the interleaved block encoding strategy to encode classical data and naturally handles the quantum data (in the form of quantum states) as input.
For each input data, the classifier determines the prediction label based on the local observable $\langle \hat{\sigma}^{z}_9\rangle$: label $0$ and label $1$ for $\langle \hat{\sigma}^{z}_9\rangle \geq0$ and $\langle \hat{\sigma}^{z}_9\rangle <0$, respectively.
}
\label{fig:18_qubit_circuit}
\end{figure*}

\begin{figure*}[htb]
\includegraphics[width=0.85\textwidth]{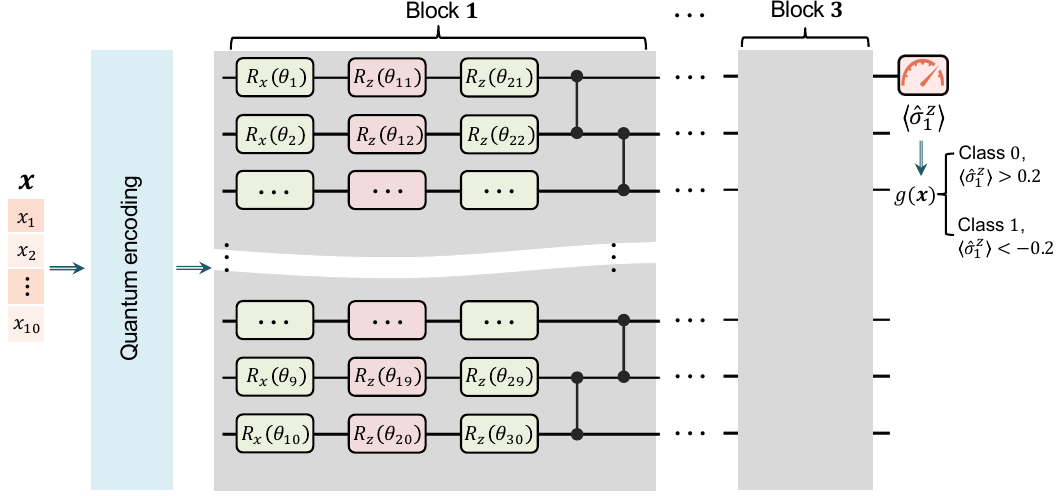}
\caption{ 
\textbf{Dataset generation for the engineered quantum task.} 
Each ten-dimensional input vector is first embedded into a quantum state via the quantum feature encoding (Supplementary Fig.~2\textbf{b}).
A variational circuit with randomly chosen parameters is then applied to the state.
The ground true label of each input image vector is generated based on the local observable $\langle \hat{\sigma}^{z}_1\rangle$. 
}
\label{fig:exp_relabel}
\end{figure*}

\bibliography{qcl}

\clearpage
\newpage 
\onecolumngrid

\newcommand{\mcol}{\multicolumn{2}{c}}
\newcommand{\cfill}{\cellcolor[HTML]{44cef6}}
\newcommand{\cfillo}{\cellcolor[HTML]{f05654}}
\setcounter{MaxMatrixCols}{10}
\hypersetup{urlcolor=blue}
\xdef\presupfigures{\arabic{figure}}
\newcommand{\figpath}{./figures}
\renewcommand{\algorithmicrequire}{\textbf{Input}}  
\renewcommand{\algorithmicensure}{\textbf{Output}} 
\setcounter{figure}{0}
\renewcommand{\theequation}{S\arabic{equation}}
\renewcommand{\thefigure}{S\arabic{figure}}
\renewcommand{\figurename}{Supplementary Fig.}

\begin{center} 
	{\large \bf Supplementary: Experimental quantum continual learning with  programmable \\
superconducting qubits}
\end{center} 

\maketitle

\section{Theoretical details}

\subsection{Variational quantum learning models}

Over the recent years, quantum computing and artificial intelligence (AI) have made dramatic progress~\cite{Arute2019Quantum,Zhong2020Quantum,Wu2021Strong,Silver2017Mastering,Jumper2021Highly,OpenAI2024GPT4}. The interplay between two promising fields gave rise to a new field called quantum machine learning~\cite{Biamonte2017Quantum,DasSarma2019Machine,Dunjko2018Machine,Cerezo2022Challenges}. 
By harnessing quantum properties such as superposition and entanglement, quantum machine learning approaches hold the potential to bring advantages
compared with their classical counterpart~\cite{Huang2021InformationTheoretic,Gao2018Quantum,Liu2021Rigorous}.
Undoubtedly, the emergent research frontier of quantum machine learning has become one of today's most rapidly growing interdisciplinary fields.
In the noisy intermediate-scale quantum (NISQ) era~\cite{Preskill2018Quantum,Bharti2022Noisy}, variational
quantum algorithms (VQA)~\cite{Cerezo2021Variational}, which exploit the hybrid quantum-classical scheme, have become a popular approach to handling classical and quantum data in practical problems~\cite{Li2022Quantum}.
As shown in Supplementary Fig.~\ref{fig:VQA}, VQA executes a parameterized quantum circuit on a quantum device and iteratively optimizes the circuit parameters through a classical optimizer to minimize a certain loss function.

In this section, we focus on classification tasks and introduce data encoding, circuit structures, and optimization strategies for variational quantum circuits.

\subsubsection{Data encoding and circuit structures} 
Before handling the original data, it is necessary to transform them into quantum states, serving as the input for quantum circuits.
Quantum data, in the form of quantum states, can be directly processed by quantum circuits. 
For classical data, various encoding circuits have been proposed to embed the original vector into the Hilbert space of qubits~\cite{Havlicek2019Supervised,Ren2022Experimental,Schuld2021Machine,Schuld2021Effect}.

In this subsection, we introduce three options for quantum encoding that were utilized in our experiments, as shown in Supplementary Fig.~\ref{fig:three_encoding}.
In Supplementary Fig.~\ref{fig:three_encoding}\textbf{a}, we show the interleaved block encoding strategy~\cite{Ren2022Experimental}, which is used to encode large-size real-life images ($16\times16$ pixels in this work).
In Supplementary Fig.~\ref{fig:three_encoding}\textbf{b}, we show the quantum circuit that implements the feature encoding proposed by Ref.~\cite{Havlicek2019Supervised}. In the main text, we adapt the feature encoding to encode the engineered quantum data.
The feature encoding maps the input data $\boldsymbol{x}$ into the quantum state $|\phi(\boldsymbol{x})\rangle$ with the form:
\begin{equation}
    |\phi(\boldsymbol{x})\rangle = U_\phi(\boldsymbol{x})\left|0^{\otimes n}\right\rangle = U_z(\boldsymbol{x}) H^{\otimes n} U_z(\boldsymbol{x}) H^{\otimes n}\left|0^{\otimes n}\right\rangle  ,
\label{eq:feature_encoding}
\end{equation}
where $H$ is is the Hadamard gate, $U_z(\boldsymbol{x})$ is a diagonal matrix in the Pauli Z-basis.
For experimental implementation, we consider $U_z(\boldsymbol{x})$ in a hardware-efficient form with only the nearest interactions:
\begin{equation}
U_z(\boldsymbol{x}) = \exp \left(-\frac{\mathrm{i}}{2}\ \left[\sum_{i=1}^D x_i \hat{\sigma}^{z}_{i} + t \sum_{\substack{<i, j>}} x_i x_j \hat{\sigma}^{z}_{i} \hat{\sigma}^{z}_{j} \right]\right) ,
\end{equation}
where the hyperparameter $t$ controls the strength of the nearest interactions and is set to be $4$ in our experiment, and $\langle i, j\rangle$ denotes the nearest qubits.
In Supplementary Fig.~\ref{fig:three_encoding}\textbf{c}, we show a rotation encoding circuit~\cite{Schuld2021Machine}. 

\begin{figure*}[t]
\centering
\includegraphics[width=0.8\textwidth]{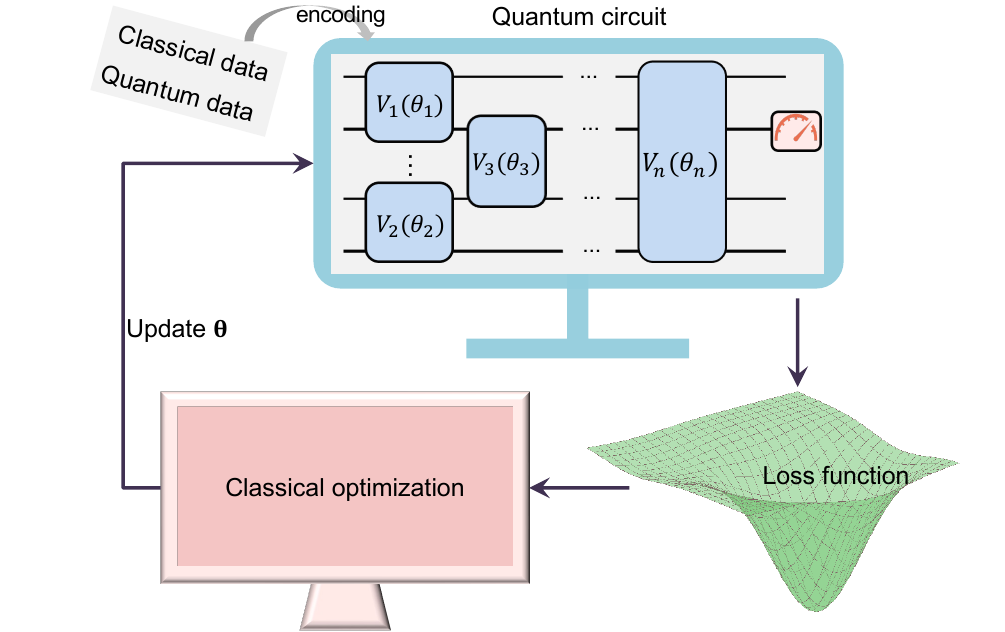}
\caption{
\textbf{Framework of variational quantum algorithms.} 
Classical or quantum data is encoded into a variational quantum circuit. The output for data samples is obtained through quantum measurements. Based on measurement outcomes, the loss function and gradient are calculated and then sent to a classical optimizer, which updates the parameters in the quantum circuit.
}
\label{fig:VQA}
\end{figure*}

\begin{figure*}[t]
\centering
\includegraphics[width=0.75\textwidth]{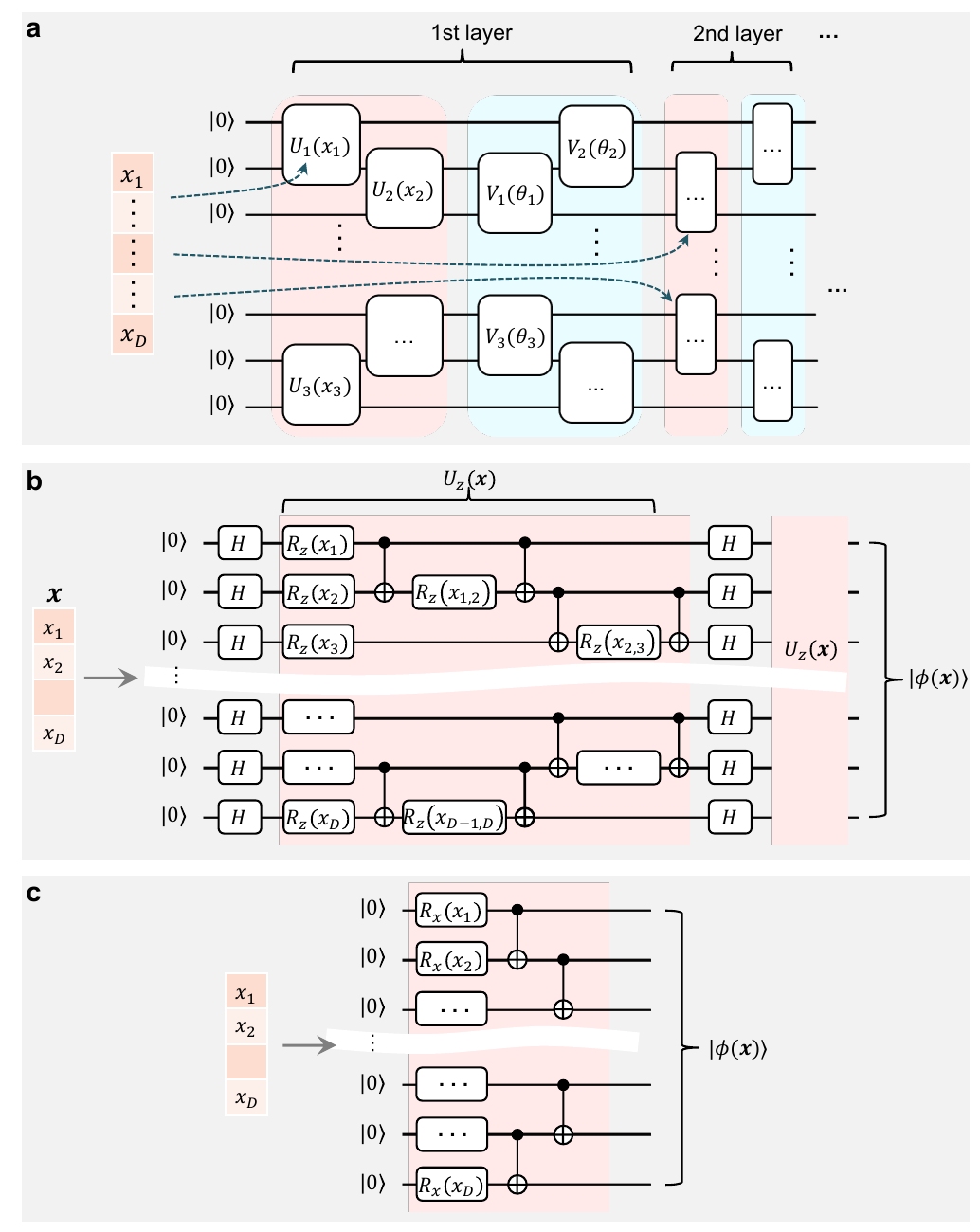}
\caption{
\textbf{Data encoding strategies.} 
\textbf{a,} Schematic illustration of the interleaved block encoding strategy. The quantum circuit alternates between encoding parts and variational parts, where the encoding part encodes the input data and the variational part is used for optimization.
\textbf{b,} Quantum circuit for implementing the feature encoding $U_{\phi}(\boldsymbol{x})$ in Equation.~(\ref{eq:feature_encoding}). Here, $x_{i,j}=t x_{i}x_{j}$.
\textbf{c,} Quantum circuit for implementing rotation encoding. Each entry of the input vector is encoded into a single-qubit state by applying a $R_x$ gate, followed by an entangling part that performs CNOT gates on adjacent qubits.
}
\label{fig:three_encoding}
\end{figure*}

After encoding the input data as quantum states, we utilize variational quantum circuits to handle them.
The circuit structure is typically designed to be hardware-efficient.
As shown in Supplementary Fig.~\ref{fig:variational}, we construct a circuit containing $p$ blocks of operations, each block including a rotation part and an entangling part. In the rotation part, each qubit undergoes three independent rotation gates $R_z$, $R_x$, and $R_z$ with specific rotation angles. These angles serve as variational parameters to be updated during the optimization process.

\begin{figure*}[t]
\centering
\includegraphics[width=0.9\textwidth]{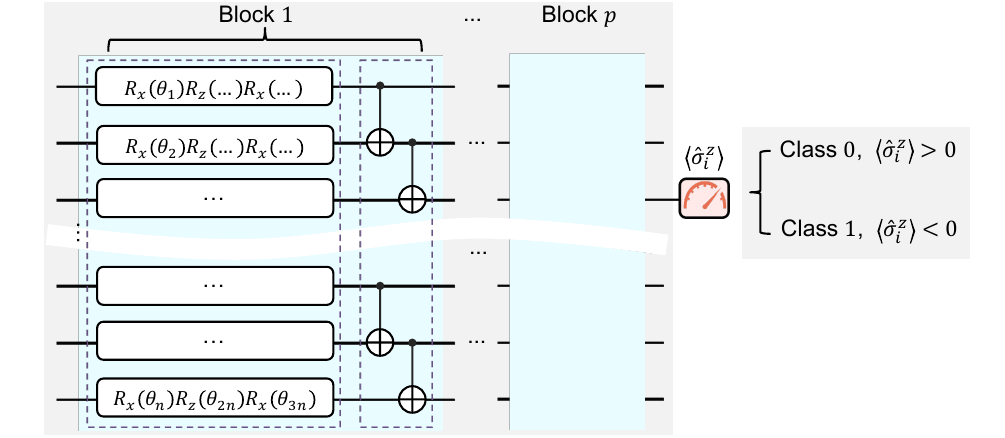}
\caption{
\textbf{A type of hardware-efficient structure for variational quantum circuits.} 
The variational quantum circuit contains $p$ blocks of operations. Each block contains a rotation part that performs three consecutive rotation gates $R_x-R_z-R_x$ on each qubit and an entangling part that performs CNOT gates on adjacent qubits. The prediction label is based on the measurement outcome $\langle\hat{\sigma}^{z}_{i}\rangle$ on $i$-th qubit.
}
\label{fig:variational}
\end{figure*}

\subsubsection{Optimization strategies} 
In our work, we mainly focus on binary classification tasks, where each input data sample is associated with a label of either $0$ or $1$.
Variational quantum circuits serve as classifiers, which can be optimized to assign correct labels to each data sample.
For each input data $\boldsymbol{x}_i$, the quantum circuit produces a corresponding output state $|\psi_i\rangle$.
As shown in Supplementary Fig.~\ref{fig:variational}, we determine the predicted label for $\boldsymbol{x}_i$ based on outcomes of projective measurements on the $m$-th qubit in the Pauli Z-basis.
The expectation values of the projector $O^{l}_{m} = |l\rangle\langle l|_m$, i.e., $\langle \psi_i | O^{l}_m |\psi_i \rangle$, represents the probability of assigning the label $l$ ($l = 0, 1$) to $\boldsymbol{x}_i$.
If $\langle \psi_i | O^{0}_m |\psi_i \rangle$ is larger than $\langle \psi_i | O^{1}_m |\psi_i \rangle$, we assign  label $0$ to $\boldsymbol{x}_i$, and vice versa.

In the optimization procedure, we use the cross-entropy loss function to measure the disparity between the predicted output and the true label, which is expressed as:
\begin{equation}
\begin{aligned}
    L(\boldsymbol{\theta}) 
    &= \frac{1}{N} \sum_{i=1}^N L\left(h\left(\boldsymbol{x}_{i} ; \boldsymbol{\theta} \right), \mathbf{a}_{i}\right) \\
    &= - \frac{1}{N} \sum_{i=1}^N (\mathbf{a}_{i}^{0}\log \mathbf{g}_{i}^0 + \mathbf{a}_{i}^{1}\log \mathbf{g}_{i}^1)  ,
\end{aligned}
\label{eq:cross_entropy}
\end{equation}
where $N$ denotes the number of training samples, $\mathbf{a}_{i} = (\mathbf{a}_{i}^{0}, \mathbf{a}_{i}^{1})$ denotes the true label of $\boldsymbol{x}_i$ in the form of one-hot encoding, $h\left(\boldsymbol{x}_{i}; \boldsymbol{\theta} \right)$ denotes the hypothesis function for the quantum circuit parameterized by $\boldsymbol{\theta}$, and
$\mathbf{g}_{i} = (\mathbf{g}_{i}^{0}, \mathbf{g}_{i}^{1})$ denotes the probability of  $\boldsymbol{x}_i$ being predicted as label $0$ or label $1$.

During the training process, we can use the stochastic gradient descent method to iteratively update the variational parameters to minimize the loss function.
Considering a single data sample $\boldsymbol{x}_i$, the derivatives $\frac{\partial L}{\partial \boldsymbol{\theta}}$ can be expressed as:
\begin{equation}
\frac{\partial L}{\partial \boldsymbol{\theta}}=-\frac{\mathbf{a}_{i}^{0}}{\mathbf{g}_{i}^{0}} \frac{\partial \mathbf{g}_{i}^{0}}{\partial \boldsymbol{\theta}}-\frac{\mathbf{a}_{i}^{1}}{\mathbf{g}_{i}^{1}} \frac{\partial \mathbf{g}_{i}^{1}}{\partial \boldsymbol{\theta}}  .
\end{equation}
When a gate with parameter $\theta$ is in the form $\exp(-\frac{i}{2}\theta P_n)$ ($P_n$ belong to the Pauli group), the derivatives of $\mathbf{g}_{i}^{l}$ can be computed via the ``parameter shift rule''~\cite{Li2017Hybrid,Mitarai2018Quantum}:
\begin{equation}
    \frac{\partial \mathbf{g}_{i}^{l}}{\partial \theta} = \frac{(\mathbf{g}_{i}^{l})^{+} - (\mathbf{g}_{i}^{l})^{-}}{2} ,
\end{equation}
where $(\mathbf{g}_{i}^{l})^{\pm}$ denotes the expectation values of $O^{l}_m$ with parameter $\theta$ being replaced by $\theta \pm \frac{\pi}{2}$.
With the obtained gradient, we can use the Adam optimizer~\cite{Kingma2017Adam} or the Nadam optimizer~\cite{Dozat2016Incorporating} to gain higher training performance in practice.

\subsection{Quantum continual learning}
In classical machine learning, artificial neural networks need to incrementally learn from dynamic streams of data, as the external environment tends to exhibit dynamic characteristics.
A major challenge in this context is known as catastrophic forgetting~\cite{McCloskey1989Catastrophic,Goodfellow2015Empirical}, where training a model on new tasks would result in a dramatic performance drop on previously learned ones.
This issue reflects the trade-off between learning plasticity and memory stability: too much emphasis on one can disrupt the other.
On the one hand, the model needs sufficient plasticity to accommodate new tasks, but large changes in the network's parameters generally lead to a dramatic performance drop on previously learned tasks.
On the other hand, maintaining stable network parameters can prevent the forgetting of old tasks but will hinder the model's ability to learn new ones.
The field of continual learning seeks to enable learning models to continually acquire knowledge from dynamic environments~\cite{Wang2024Comprehensive,Ditzler2015Learning,chen2022lifelong}.
It achieves this by effectively balancing memory stability with learning plasticity.

With the rapid development of quantum machine learning, quantum computers also need to handle dynamic streams of data in the real world.
Yet, most existing quantum learning models are designed for isolated tasks. 
Enabling quantum models to incrementally accumulate knowledge from dynamic environments is crucial for achieving the long-term goal of quantum artificial general intelligence.
More recently, it has been shown theoretically that quantum neural networks also suffer from catastrophic forgetting in continual learning scenarios~\cite{Jiang2022Quantum}.
A naive baseline approach to mitigate catastrophic forgetting is to retrain quantum learning models using training samples from all previously learned tasks. 
However, this approach is resource-expensive in terms of both training costs and data storage.
Additionally, unlike classical data, quantum data (in the form of quantum states) cannot be cloned perfectly~\cite{Wootters1982Single}, adding a unique challenge to retraining.
Therefore, to address the above-mentioned problem, it is crucial to develop effective quantum continual learning methods without the need for old training samples.

\subsubsection{Elastic weight consolidation}
To mitigate catastrophic forgetting, a variety of strategies have been proposed, including regularization-based approaches~\cite{Zenke2017Continual,Kirkpatrick2017Overcoming}, optimization-based approaches~\cite{Chaudhry2018Efficient,Bennani2020Generalisation,Farajtabar2020Orthogonal}, architecture-based approaches~\cite{Rusu2022Progressive}, and among others.
Our experiments have demonstrated that variational quantum circuit classifiers, similar to their classical counterparts, suffer from catastrophic forgetting. 
In the main text, we have demonstrated the Elastic Weight Consolidation (EWC) method, which falls under regularization-based approaches, can effectively mitigate catastrophic forgetting in continual learning scenarios.
In this section, we provide a detailed discussion on the application of EWC in quantum continual learning scenarios.

Recent studies suggest that animals preserve memories of previously learned knowledge by strengthening specific excitatory synapses based on past experiences. Inspired by this, EWC preserves memories of previous tasks by selectively penalizing the variation of each parameter according to its importance.

Let's consider a learning model with parameter $\boldsymbol{\theta}$ incrementally learning two sequential tasks $\mathcal{T}_1$ and $\mathcal{T}_2$, using the corresponding training dataset $D_1$ and $D_2$, which are drawn from distributions $\mathcal{D}_1$ and $\mathcal{D}_2$ respectively.
The learning process consists of two stages: in the first stage, the model is trained on $\mathcal{D}_1$ to learn $\mathcal{T}_1$, while in the second stage, the model does not have access to $\mathcal{D}_1$ and is only trained on $\mathcal{D}_2$ to learn $\mathcal{T}_2$.

From the probabilistic perspective, the goal is to find parameter $\boldsymbol{\theta}$ that maximizes the posterior probability $p(\boldsymbol{\theta} | \mathcal{D}_1,\mathcal{D}_2)$.
Assuming $\mathcal{D}_1$ and $\mathcal{D}_2$ are independent of each other, the posterior can be computed with Bayes' rule:
\begin{equation}
\begin{aligned}
p(\boldsymbol{\theta} | \mathcal{D}_1,\mathcal{D}_2) & =\frac{p(\boldsymbol{\theta}, \mathcal{D}_1,\mathcal{D}_2)}{p(\mathcal{D}_1, \mathcal{D}_2)} \\
& =\frac{p\left(\boldsymbol{\theta}, \mathcal{D}_2 | \mathcal{D}_1\right) p\left(\mathcal{D}_1\right)}{p\left(\mathcal{D}_2 | \mathcal{D}_1\right) p\left(\mathcal{D}_1\right)} \\
& =\frac{p\left(\boldsymbol{\theta}, \mathcal{D}_2 | \mathcal{D}_1\right)}{p\left(\mathcal{D}_2\right)} \\
& =\frac{p\left(\mathcal{D}_2 \mid \boldsymbol{\theta}, \mathcal{D}_1\right) p\left(\boldsymbol{\theta} \mid \mathcal{D}_1\right)}{p\left(\mathcal{D}_2\right)} \\
& =\frac{p\left(\mathcal{D}_2 | \boldsymbol{\theta}\right) p\left(\boldsymbol{\theta} | \mathcal{D}_1\right)}{p\left(\mathcal{D}_2\right)} .
\end{aligned}
\end{equation}
By taking the logarithm of both sides of the equation, we obtain:
\begin{equation}
\log p(\boldsymbol{\theta} | \mathcal{D}_1,\mathcal{D}_2) = \log p\left(\mathcal{D}_2 | \boldsymbol{\theta}\right) + \log p\left(\boldsymbol{\theta} | \mathcal{D}_1\right) - \log p\left(\mathcal{D}_2 \right).
\label{eq:posterior}
\end{equation}
Here, $\log(\mathcal{D}_2 | \boldsymbol{\theta})$ represents the negative of the loss function $-L_2(\boldsymbol{\theta})$ for task $\mathcal{T}_2$.
The constant term $\log p\left(\mathcal{D}_2 \right)$ can be omitted.
The posterior $p\left(\boldsymbol{\theta} | \mathcal{D}_1\right)$ containing all the information about $\mathcal{T}_1$, is generally intractable. 
EWC approximates it as a multivariate Gaussian distribution centered at the maximum value $\boldsymbol{\theta}_1^{\star}$:
\begin{equation}
\log p\left(\boldsymbol{\theta} | \mathcal{D}_1\right) = \log p\left(\boldsymbol{\theta}_1^{\star} | \mathcal{D}_1\right) + \left.\frac{\partial \log p\left(\boldsymbol{\theta} | \mathcal{D}_1\right)}{\partial \boldsymbol{\theta}} \right|_{\boldsymbol{\theta}=\boldsymbol{\theta}_1^{\star} }(\boldsymbol{\theta} - \boldsymbol{\theta}_1^{\star} ) + \frac{1}{2}(\boldsymbol{\theta} - \boldsymbol{\theta}_1^{\star} )^T H_{\boldsymbol{\theta}_1^{\star}} (\boldsymbol{\theta} - \boldsymbol{\theta}_1^{\star} ) ,
\end{equation}
where $H_{\boldsymbol{\theta}_1^{\star}}$ is the Hessian matrix. As $\boldsymbol{\theta}_1^{\star}$ maximizes $p\left(\boldsymbol{\theta}, T_1\right)$, the first-order Taylor expansion can be omitted.
Thus, Equation~(\ref{eq:posterior}) can be written as:
\begin{equation}
    \log p(\boldsymbol{\theta} | \mathcal{D}_1,\mathcal{D}_2) = - L_2(\boldsymbol{\theta}) + \frac{1}{2}(\boldsymbol{\theta} - \boldsymbol{\theta}_1^{\star} )^T H_{\boldsymbol{\theta}_1^{\star}} (\boldsymbol{\theta} - \boldsymbol{\theta}_1^{\star} )  .
\end{equation}
To maximize $p(\boldsymbol{\theta} | \mathcal{D}_1,\mathcal{D}_2)$, the loss function in the second learning stage can be defined as:
\begin{equation}
L_2^{\text{EWC}}(\boldsymbol{\theta}) = L_2(\boldsymbol{\theta}) - \frac{1}{2}(\boldsymbol{\theta} - \boldsymbol{\theta}_1^{\star} )^T H_{\boldsymbol{\theta}_1^{\star}} (\boldsymbol{\theta} - \boldsymbol{\theta}_1^{\star} ) .
\end{equation}
Here, we note that the negative expected Hessian matrix $H_{\boldsymbol{\theta}_1^{\star}}$ is equal to the Fisher information matrix $\boldsymbol{F}_1(\boldsymbol{\theta}_1^{\star})$, since:
\begin{equation}
\begin{aligned}
\mathbb{E}_{p\left( \mathcal{D}_1 | \boldsymbol{\theta} \right)}[H_{\boldsymbol{\theta}}] &= \mathbb{E}_{p\left( \mathcal{D}_1 | \boldsymbol{\theta} \right)} \left[\frac{\partial^2 \log p\left(\boldsymbol{\theta} | D_1\right)}{\partial \boldsymbol{\theta}^2} \right] \\
&= \mathbb{E}_{p\left( \mathcal{D}_1 | \boldsymbol{\theta}  \right)} \left[\frac{\partial^2 \log p\left( \mathcal{D}_1 | \boldsymbol{\theta} \right)}{\partial \boldsymbol{\theta}^2} \right] + \mathbb{E}_{p\left( \mathcal{D}_1 | \boldsymbol{\theta}  \right)} \left[\frac{\partial^2 \log p\left( \boldsymbol{\theta} \right)}{\partial \boldsymbol{\theta}^2} \right]
\\
&\approx \mathbb{E}_{p\left( \mathcal{D}_1 | \boldsymbol{\theta}  \right)} \left[\frac{\partial^2 \log p\left( \mathcal{D}_1 | \boldsymbol{\theta} \right)}{\partial \boldsymbol{\theta}^2} \right] \\
&= \mathbb{E}_{p\left( \mathcal{D}_1 | \boldsymbol{\theta} \right)} \Bigg\{\frac{\frac{\partial^2}{\partial \boldsymbol{\theta}^2} p\left(\mathcal{D}_1 | \boldsymbol{\theta}\right)}{p\left(\mathcal{D}_1 | \boldsymbol{\theta}\right)} -\left( \frac{\frac{\partial}{\partial \boldsymbol{\theta}} p\left(\mathcal{D}_1 | \boldsymbol{\theta}\right) }{p\left(\mathcal{D}_1 | \boldsymbol{\theta}\right)}\right)\left( \frac{\frac{\partial}{\partial \boldsymbol{\theta}} p\left(\mathcal{D}_1 | \boldsymbol{\theta}\right) }{p\left(\mathcal{D}_1 | \boldsymbol{\theta}\right)}\right)^\top \Bigg\} \\
&= \frac{\partial^2}{\partial\boldsymbol{\theta}^2} \int p\left(\mathcal{D}_1 | \boldsymbol{\theta}\right) d \mathcal{D}_1 - \mathbb{E}_{p\left(\mathcal{D}_1 | \boldsymbol{\theta}\right)} \Bigg[\Big(\frac{\partial}{\partial \boldsymbol{\theta}} \log p\left(\mathcal{D}_1 | \boldsymbol{\theta}\right) \Big) \Big(\frac{\partial}{\partial \boldsymbol{\theta}} \log p\left(\mathcal{D}_1 | \boldsymbol{\theta}\right) \Big)^\top \Bigg] \\
&= - \boldsymbol{F}_{1}(\boldsymbol{\theta})
\end{aligned}    
\end{equation}
Here, we approximate our calculations by omitting the prior distribution $p(\boldsymbol{\theta})$ on $\boldsymbol{\theta}$. Thus, $\mathcal{L}_2(\boldsymbol{\theta}) $ can be expressed as:
\begin{equation}
L_2^{\text{EWC}}(\boldsymbol{\theta}) = L_2(\boldsymbol{\theta}) + \frac{\lambda}{2}(\boldsymbol{\theta} - \boldsymbol{\theta}_1^{\star} )^T \boldsymbol{F}_1 (\boldsymbol{\theta} - \boldsymbol{\theta}_1^{\star} ) .
\end{equation}
Here, we use a hyperparameter $\lambda$ to control the regularization strength. To simplify the computation, we only consider the diagonal term of $\boldsymbol{F}_1$, further simplifying $\mathcal{L}_2(\boldsymbol{\theta})$ to:
\begin{equation}
L_2^{\text{EWC}}(\boldsymbol{\theta}) = L_2(\boldsymbol{\theta}) +  \frac{\lambda}{2} \sum_i F_{1,i} (\boldsymbol{\theta}_i - \boldsymbol{\theta}^{\star}_{1,i} )^2,
\label{eq:ewc}
\end{equation}
where $F_{1,i}$ is the $i$-th diagonal element of $\boldsymbol{F}_1$ at the optimal value $\boldsymbol{\theta}_1^{\star}$ for task $\mathcal{T}_1$.

In practice, $p\left( \mathcal{D}_1 | \boldsymbol{\theta} \right)$ can be evaluated on the training samples from $\mathcal{T}_1$, so $F_{1,i}$ can be estimated as:
\begin{equation}
F_{1,i} = \frac{1}{\left|D_1\right|} \sum_{(x,y_{\boldsymbol{x}}) \in D_1} \left(\left.\frac{\partial \log p
\left(y=y_{\boldsymbol{x}} | \boldsymbol{\theta} \right)}{\partial \boldsymbol{\theta}_i}\right|_{\boldsymbol{\theta}=\boldsymbol{\theta}_1^{\star}}\right)^2 ,
\end{equation}
where $p\left(y=y_{\boldsymbol{x}} | \boldsymbol{\theta} \right)$ is the probability that the prediction label $y$ for input data $\boldsymbol{x}$ the true label match the true label $y_{\boldsymbol{x}}$. This probability is determined by the learning model parameterized by $\boldsymbol{\theta}$.
A larger value of $F_{1,i}$ indicates greater importance for the parameter $\boldsymbol{\theta}_{i}$, leading to a greater penalty for deviations of $\boldsymbol{\theta}_{i}$ from $\boldsymbol{\theta}^{\star}_{1,i}$ in the regularization term of Equation~(\ref{eq:ewc}).

We can generalize the above discussion about a continual learning scenario involving two tasks to a scenario involving $T$ tasks, denoted as $\mathcal{T}_t$ ($t=1,2,\ldots,T$). 
In the scenario, the learning process consists of $T$ stages. In the $k$-th stage, $L_k^{\text{EWC}}(\boldsymbol{\theta})$ can be expressed as:
\begin{equation}
L_k^{\text{EWC}}(\boldsymbol{\theta})= L_k(\boldsymbol{\theta}) + \sum_{t=1}^{k-1} \frac{\lambda_{k,t}}{2} \sum_i F_{t,i}\left(\boldsymbol{\theta}_i-\boldsymbol{\theta}^{\star}_{t,i} \right)^2  ,
\label{eq:loss_ewc}
\end{equation}
where $\lambda_{k,t}$ controls the regularization strength for $\mathcal{T}_t$ in the $k$-th stage, and $\boldsymbol{\theta}^{\star}_{t}$ is the parameter obtained after $t$-th stage.
The Fisher information $F_{t,i}$ estimates the importance of the $i$-th parameter for $\mathcal{T}_{t}$.

\begin{algorithm}[H]
\caption{Quantum continual learning with the EWC method}
\label{algorithm 2}
\begin{algorithmic}
\REQUIRE A sequence of tasks $\mathcal{T}_1$, $\mathcal{T}_2$, $\ldots$, the variational quantum circuit classifier with parameters $\boldsymbol{\theta}$, iteration steps $S$, batch size $n_b$, learning rate $\eta$, $\lambda_{k,t}$ which controls the regularization strength for $\mathcal{T}_t$ in the $k$-th stage
\ENSURE The optimal parameter $\boldsymbol{\theta^{\star}}$ 
\STATE \textbf{Initialize} Random initial parameters $\boldsymbol{\theta}$
\FOR{ task $k=1,2,\ldots$}
\FOR{ $s\in [S]$}
\STATE Randomly choose $n_b$ samples from the training dataset for $\mathcal{T}_k$, and then calculate the gradient of the loss function over the training batch: $\boldsymbol{G} = 
\nabla_{\boldsymbol{\theta}} \Big[ L_k(\boldsymbol{\theta}) + \sum_{t=1}^{k-1} \frac{\lambda_{k,t}}{2} \sum_i F_{t,i}\left(\boldsymbol{\theta}_i-\boldsymbol{\theta}^{\star}_{t,i} \right)^2 \Big]$
\STATE  $\boldsymbol{\theta} \leftarrow f_{\text{Nadam}}(\boldsymbol{\theta}, \eta, \boldsymbol{G}) $
\ENDFOR
\STATE $\boldsymbol{\theta}_k^{\star} \leftarrow \boldsymbol{\theta}$
\STATE  Calculate the Fisher information for each parameter: 
\begin{equation}
F_{k,i} = \frac{1}{\left|D_k\right|} \sum_{(x,y_{\boldsymbol{x}}) \in D_k} \left(\left.\frac{\partial \log p
\left(y=y_{\boldsymbol{x}} | \boldsymbol{\theta} \right)}{\partial \theta_i}\right|_{\boldsymbol{\theta}=\boldsymbol{\theta}_k^{\star}}\right)^2 
\end{equation}
\ENDFOR
\end{algorithmic}
\end{algorithm}

\begin{figure}[t]
\includegraphics[width=0.7\textwidth]{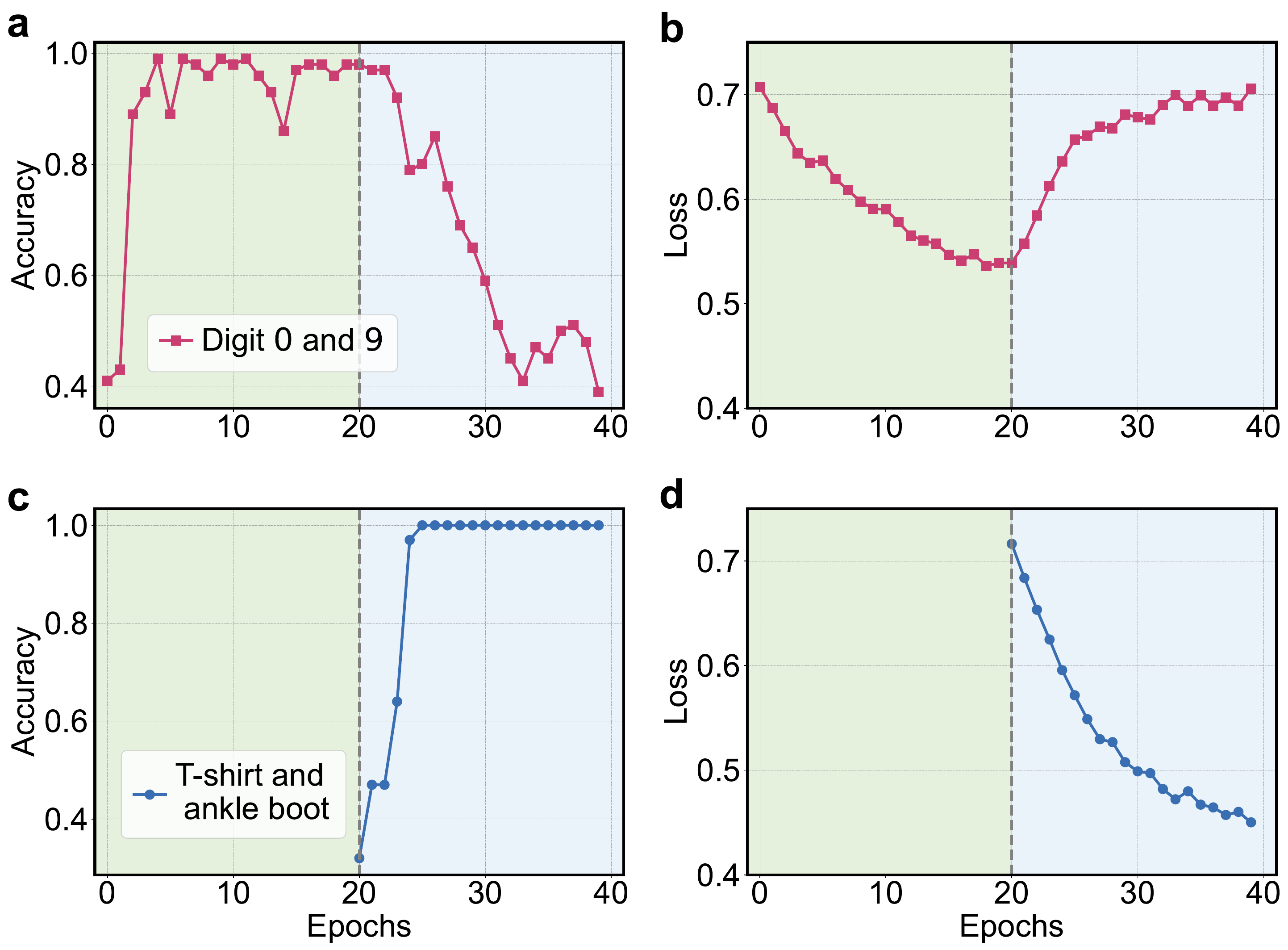}
\caption{\textbf{Numerical results for demonstrating catastrophic forgetting.}
\textbf{a and b,} Accuracy and loss function on the test dataset for $\mathcal{T}_1$ at each epoch during the first and second learning stages.
\textbf{c and d,} Accuracy and loss function on the test dataset for $\mathcal{T}_2$ as functions of training epochs during the second stage.
}
\label{fig:forgetting_numerical}
\end{figure}

\begin{figure}[t]
\includegraphics[width=0.7\textwidth]{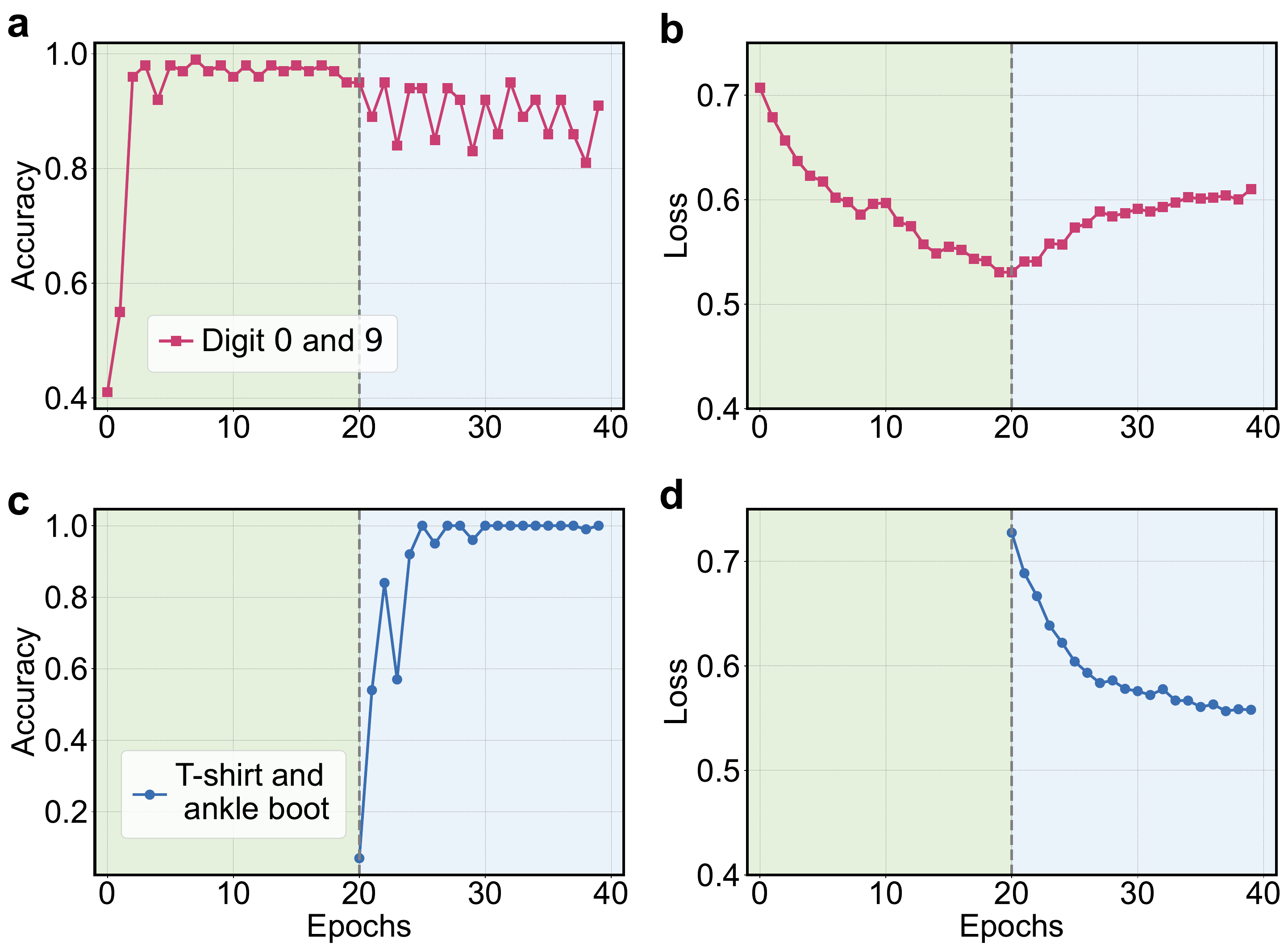}
\caption{
\textbf{Numerical results for demonstrating the EWC method in mitigating catastrophic forgetting.}
\textbf{a and b,} Prediction accuracy and loss for $\mathcal{T}_1$ at each training epoch during the incremental learning process.
\textbf{c and d,} Prediction accuracy and loss for $\mathcal{T}_2$ at each epoch during the second stage.
}
\label{fig:ewc_numerical}
\end{figure}

\begin{figure}
\includegraphics[width=0.7\textwidth]{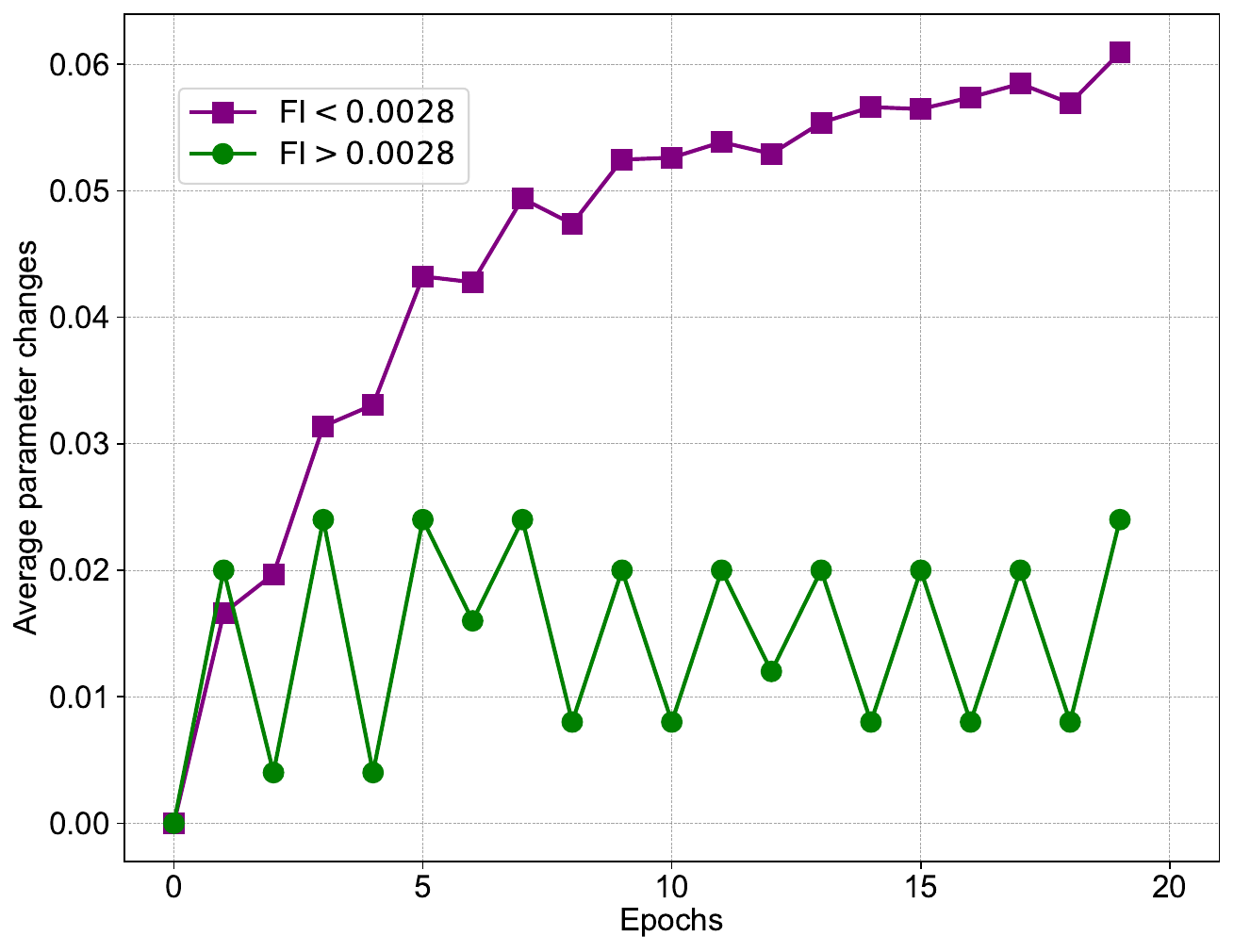}
\caption{\textbf{Average parameter changes during the second stage for learning $\mathcal{T}_2$.}
We divide all parameters into two groups based on $F_{1,i}>0.028$ or $F_{1,i}<0.028$. We plot the average parameter change for each group as a function of training epochs during the second learning stage.
}
\label{fig:average_parameter_changes}
\end{figure}

\subsubsection{Numerical simulations and benchmarks}
In this subsection, we focus on variational quantum learning models and conduct classical simulations to benchmark the performance of the EWC method in mitigating catastrophic forgetting.

We consider a learning model incrementally learning two classification tasks $\mathcal{T}_1$, $\mathcal{T}_2$ in sequence.
$\mathcal{T}_1$ concerns the classification of clothing images labeled as ``T-shirt'' and ``ankle boot'' from the Fashion-MNIST dataset.
$\mathcal{T}_2$ is about classifying medical magnetic resonance imaging (MRI) scans labeled as ``hand'' and ``breast''.
For each task, we randomly select $500$ training samples and $100$ test samples from the corresponding dataset.
The classical data ($\mathcal{T}_1$ and $\mathcal{T}_2$) is encoded into quantum circuits using the interleaved block-encoding strategy.
In our numerical simulations, we construct a $10$-qubit variational quantum classifier, with the structure exhibited in Supplementary Fig.~\ref{fig:variational}. The quantum classifier contains $9$ blocks of operations, with a total of $270$ variational parameters.

To demonstrate catastrophic forgetting, we train the quantum classifier with the cross-entropy loss function defined in Equation~(\ref{eq:cross_entropy}) for the two tasks.
During the training process, we set the learning rate to $0.02$ assisted by the Adam optimizer, and the batch size to $25$.
Our numerical results are displayed in Supplementary Fig.~\ref{fig:forgetting_numerical}. The learning process consists of two learning stages. In the first learning stage, the quantum classifier is trained to learn $\mathcal{T}_1$. After $20$ epochs, the quantum classifier achieves a $98\%$ accuracy and $54\%$ loss on the test dataset for $\mathcal{T}_1$.
In the second stage, the quantum classifier is retrained to learn $\mathcal{T}_2$. After $20$ training epochs, the prediction accuracy on $\mathcal{T}_2$ reaches $100\%$. However, after the second stage the accuracy on $\mathcal{T}_1$ dramatically drops to $39\%$.
These numerical results clearly manifest the catastrophic forgetting phenomena for quantum learning.

We also conduct numerical simulations to show that the catastrophic forgetting problem can be effectively overcome with the EWC method. 
To this end, we train the quantum classifier with the modified loss function as defined in Equation~(\ref{eq:loss_ewc}) in the second stage for learning $\mathcal{T}_2$. The regularization strength $\lambda$ is set as $200$.
The numerical results are shown in Supplementary Fig.~\ref{fig:ewc_numerical}. 
We observe that after the second learning stage, the prediction accuracy for $\mathcal{T}_2$ reaches $100\%$ while the accuracy for $\mathcal{T}_1$ still maintains above $90\%$.
The results are in sharp contrast to the case without applying the EWC strategy, which confirms that the learned knowledge for $\mathcal{T}_1$ is effectively preserved with EWC.
We can observe the competitive relationship between learning plasticity for the new task and memory stability for the previous task from the loss curves in Supplementary Fig.~\ref{fig:forgetting_numerical} and Supplementary Fig.~\ref{fig:ewc_numerical}.
In the second learning stage, the increase in the loss curve for learning $\mathcal{T}_1$ in Supplementary Fig.~\ref{fig:ewc_numerical}\textbf{b} is relatively smaller than that in Supplementary Fig.~\ref{fig:forgetting_numerical}\textbf{b}, and the decrease in the loss curve for learning $\mathcal{T}_2$ in Supplementary Fig.~\ref{fig:ewc_numerical}\textbf{d} is also smaller compared to that in Supplementary Fig.~\ref{fig:forgetting_numerical}\textbf{d}.

To understand the EWC mechanism, we analyze the average parameter changes in scenarios with EWC. 
According to Equation~(\ref{eq:ewc}), parameters with large Fisher information will lead to a significant increase in the loss function if they deviate from their optimal values for previous tasks.
Therefore, to minimize the increase in the loss function, the parameters with larger Fisher information tend to undergo smaller adjustments when learning a new task.
We verify this in the numerical simulation. After learning $\mathcal{T}_1$, we divide all parameters into two groups. The first group contains $10$ parameters with $F_{1,i}$ values larger than $0.028$, and the second group contains the other $260$ parameters with $F_{1,i}$ values less than $0.0028$. We plot the average parameter change for each group during the second learning stage for $\mathcal{T}_2$.
The results are shown in Supplementary Fig.~\ref{fig:average_parameter_changes}. We observe that when applying EWC, the parameters with larger Fisher information experience smaller changes during the training process.

\section{Experimental details}

\subsection{Device information}
As shown in Fig.~1(a) in the main text, the superconducting quantum processor used in our experiments has $121$ frequency-tunable transmon qubits encapsulated in an $11$ $\times$ $11$ square lattice and $220$ tunable couplers between adjacent qubits. 
The maximum resonance frequencies of the qubits and couplers are around $4.8$~GHz and $9.0$~GHz, respectively. Each qubit has an individual XY-control line to carry out single-qubit rotations and an individual Z-control line to tune its resonance frequency. 
Each coupler also has an individual Z-control line, which can be used to modulate the effective coupling strength between two adjacent qubits dynamically with a range of up to $-25$ MHz. 
In the experiment, we select a chain of $18$ qubits to construct the variational quantum circuit classifier. 
As shown in Supplementary Fig.~\ref{fig:device parameters}\textbf{a}, the idle frequencies of the $18$ qubits distribute from $3.8$~GHz to $4.4$~GHz. 
The typical energy relaxation time $T_1$ and spin-echo dephasing time $T_2^{\text{SE}}$ at idle frequencies are shown in Supplementary Fig.~\ref{fig:device parameters}\textbf{b}, with average values of $83.2~\mu$s and $21.4~\mu$s, respectively.
Additionally, we use simultaneous cross-entropy benchmarking (XEB) to characterize the performance of the single- and two-qubit gates.
The operation errors including single-qubit gate error, two-qubit gate error, and readout error are shown in Supplementary Fig.~\ref{fig:device parameters}\textbf{c}, with average errors of $5.2 \times 10^{-4}$ and $3.9 \times 10^{-3}$ for the single-qubit and two-qubit gates, respectively.

\subsection{Continual learning of three tasks}

\subsubsection{Experimental quantum circuit classifier}
In the main text, we implement a quantum classifier with $18$ superconducting qubits to sequentially learn three tasks. 
As shown in Extended Data Fig.~1, we display the experimental circuit for data encoding and continual learning for the three tasks.
The experimental circuit consists of four blocks of operations, where each block includes three rotation gates $R_x(\theta)$, $R_z(\theta)$, $R_x(\theta)$ with independent angles on each qubit, followed by two layers of CNOT gates on adjacent qubit pairs.
Here, $R_x(\theta)$ and $R_z(\theta)$ represent the rotation gates that rotate the single-qubit quantum state by angle $\theta$ around $x$- and $z$-axis, respectively.
In our experiment, $R_x(\theta)$ are realized by applying $30$~ns microwave pulses with a full-width half-maximum of $15$~ns and shaped with DRAG 
technique~\cite{Song201710Qubit} and $R_z(\theta)$ are realized by virtual-Z gates~\cite{PhysRevA.96.022330}. 
The CNOT gate is implemented by applying a controlled $\pi$-phase (CZ) gate sandwiched between two Hadamard gates.
The CZ gate is realized by bringing $|11\rangle$ and $|02\rangle$ ($|20\rangle$) states of the qubit pairs into near resonance and steering the resonant frequency of the coupler to activate the coupling for a specific duration of $30$~ns, with the detailed calibration procedure given in Ref.~\cite{Ren2022Experimental}.

\begin{figure}[t]
\includegraphics[width=0.9\textwidth]{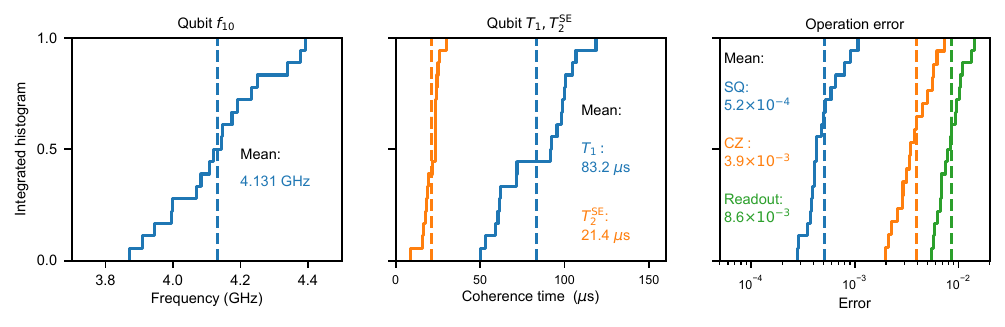}
\caption{ 
\textbf{Characteristic parameters of the superconducting quantum processor.}
\textbf{a}, Idle frequencies of the $18$ qubits used on the quantum processor. 
\textbf{b}, Relaxing times $T_1$ and spin-echo dephasing times $T_2^{
\text{SE}}$ of the $18$ qubits measured at the corresponding idle frequencies. 
\textbf{c}, Pauli errors of the simultaneous single- and two-qubit gates, and readout errors of the 18 qubits.
}
\label{fig:device parameters}
\end{figure}

\subsubsection{Encoding of the classical data}
In the main text, we experimentally implement the quantum classifier to classify classical images from the Fashion-MNIST dataset and the medical magnetic resonance imaging dataset.
For each classical image, we first represent it as a vector, with each entry corresponding to a pixel. We then use the interleaved block encoding method to embed the input vector, as illustrated in Extended Data Fig.~1.
The experimental circuit contains four blocks of operations, with a total of $216$ rotation angles.
To realize the alternation between encoding parts and variational parts, we assign the $i$-th rotation gate with an angle of $cx_i+\theta_i$, where $x_i$ and $\theta_i$ represent the $i$-th entry of the input data vector $\boldsymbol{x}$ and the variational parameter $\boldsymbol{\theta}$, respectively, and $c$ is a hyperparameter.
The alternation is thus realized since a rotation gate with combined angles can be equivalently represented by two separate rotation gates with angles $cx_i$ and $\theta_i$, respectively. In experiments, we choose $c=2$.

\subsubsection{Variational preparation of the quantum data}

In the main text, we experimentally implement the quantum classifier to recognize quantum states in antiferromagnetic (ATF) and symmetry-protected topological (SPT) phases~\cite{Smacchia2011Statistical}.
We consider the cluster-Ising model under open boundary conditions, with the following Hamiltonian:
\begin{equation}
H(h)=-\sum_{j=2}^{N-1} \hat{\sigma}_{j-1}^x \hat{\sigma}_j^z \hat{\sigma}_{j+1}^x + h \sum_{j=1}^{N-1} \hat{\sigma}_j^y \hat{\sigma}_{j+1}^y, 
\label{eq:cluster_ising}
\end{equation}
where $N=18$ denotes the total number of qubits, $h$ describes the strength of the nearest-neighbor interactions, and $\hat{\sigma}_{k}^l$ ($l=x, y, z$) is the Pauli operator on the $k$-th qubit.
This model is exactly solvable and the ground states of Equation~(\ref{eq:cluster_ising}) undergo a transition from an SPT cluster phase to an ATF phase at $h=1$.
The SPT phase is distinguished from the ATF phase by the string order parameter~\cite{Smacchia2011Statistical}:
\begin{equation}
    \langle O_z\rangle = (-1)^N\left\langle\hat{\sigma}_1^x \hat{\sigma}_2^y\left(\prod_{k=3}^{N-2} \hat{\sigma}_k^z\right) \hat{\sigma}_{N-1}^y \hat{\sigma}_N^x\right\rangle .
\end{equation}
In the thermodynamic limit $N\rightarrow \infty$, the string order parameter $\langle O_z\rangle$ can be computed as~\cite{Smacchia2011Statistical}:
\begin{equation}
\langle O_z\rangle = \begin{cases}\left(1 - h^2\right)^{3/4}, & \text { for } h<1 \\ 0 , & \text { for } h>1\end{cases} .
\end{equation}
We randomly sample multiple different values of $h$ from the range 
$[0,0.5]\cup[2.5,3]$ and prepare the ground states of their corresponding Hamiltonian. These quantum states, along with their labels (the SPT or ATF phase), form a quantum dataset.

In our experiments, we prepare approximate ground states of $H(h)$ by executing a variational circuit $U(\boldsymbol{\alpha})$ with the structure shown in Supplementary Fig.~\ref{fig:state_prepare}. The variational circuit consists of five blocks of operations, with each block containing three layers of single-qubit gates followed by two layers of CZ gates.
To determine the variational parameters $\boldsymbol{\alpha}$ for ground states of $H(h)$, we train the circuit on a classical computer to minimize the energy expectation value $\left\langle H(h)\right\rangle$ for the output states $\left|\varphi(\boldsymbol{\alpha})\right\rangle = U(\boldsymbol{\alpha}) |0\rangle$.
In the simulation, for each $H(h)$, we train the variational circuit with randomly selected initial parameters, and iteratively adjust these parameters using the gradient descent method until $\langle H(h)\rangle$ converges.
We then experimentally implement the variational circuit with the optimal parameters $\boldsymbol{\alpha^*}$ obtained in classical simulation.
To characterize our quantum state preparation, we measure the string order parameter $\langle O_z\rangle$ for these prepared states.
As shown in Supplementary Fig.~\ref{fig:Oz}, nonvanishing values of $\langle O_z\rangle$ are observed for states in the SPT phase, while $\langle O_z\rangle$ is near zero for states in the ATF phase.
This is consistent with the theoretical values for $\langle O_z\rangle$ in different phases, despite some deviations due to experimental imperfections.
We prepare a training set with a total of $500$ different states and a test set with $100$ different states.

\begin{figure}[t]
\includegraphics[width=0.7\textwidth]{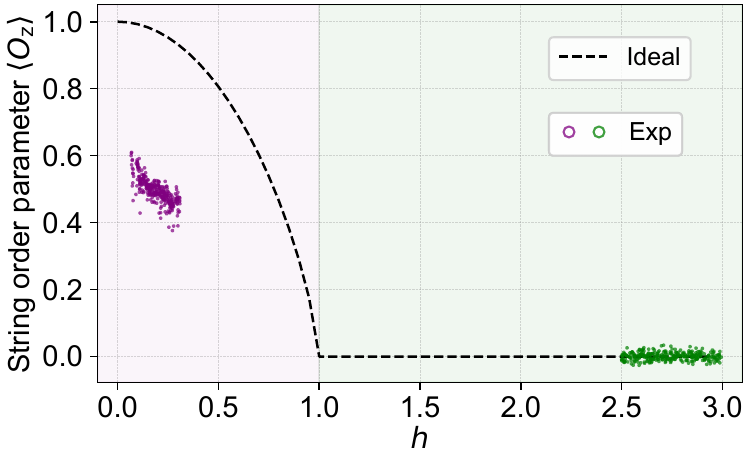}
\caption{ 
\textbf{Experimentally measured string order parameter $\boldsymbol{\langle O_z\rangle}$ for quantum states in antiferromagnetic and symmetry-protected topological phases.}
Purple and green dots represent the experimentally measured $\langle O_z\rangle$ for quantum states in the SPT phase and ATF phase, respectively.
The black dashed line plots the ideal value of $\langle O_z\rangle$ as a function of $h$.
}
\label{fig:Oz}
\end{figure}
\begin{figure}[t]
\includegraphics[width=1\textwidth]{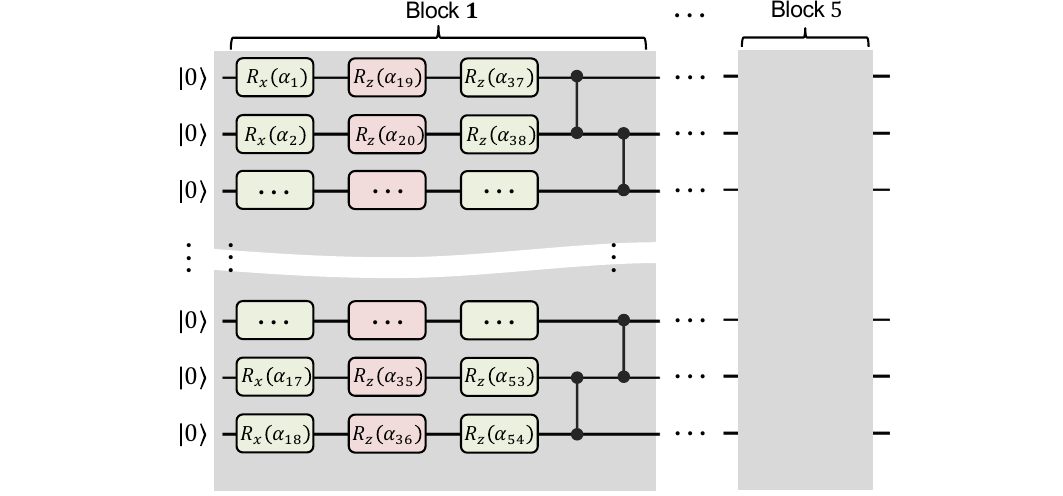}
\caption{ 
\textbf{Experimental quantum circuit with $\bold{18}$ superconducting qubits for preparing the ground states of cluster-Ising Hamiltonians.} 
The rotating angles $\alpha_i$ are obtained by numerical variational optimization.   
}
\label{fig:state_prepare}
\end{figure}

\end{document}